\documentclass[a4paper,iop,twocolumn,numberedappendix,appendixfloats]{emulateapj}
\usepackage[breaklinks,colorlinks=true,linkcolor=blue,anchorcolor=blue,citecolor=blue,urlcolor=blue,draft=false]{hyperref}
\usepackage{color}
\usepackage{graphicx}
\usepackage{amsmath} 
\usepackage{amssymb}
\usepackage{mathrsfs}
\usepackage{placeins}
\usepackage{times}
\usepackage{xcolor}
\usepackage{xspace}

\hypersetup{pdfauthor={Keiichi Umetsu},pdftitle={Ensemble Lensing}}
\usepackage{natbib}
 
\newcommand{\simgt}{\lower.5ex\hbox{$\; \buildrel > \over \sim \;$}}
\newcommand{\simlt}{\lower.5ex\hbox{$\; \buildrel < \over \sim \;$}}

\newcommand{\Mvir}{M_\mathrm{vir}}

\newcommand{\rvir}{r_\mathrm{vir}}
\newcommand{\Rsp}{R_\mathrm{sp}^\mathrm{3D}}

\newcommand{\kpch}{h^{-1}\,\mathrm{kpc}}
\newcommand{\Mpch}{h^{-1}\,\mathrm{Mpc}}
\newcommand{\Msunh}{h^{-1}\,M_\odot}

\newcommand{\argmin}{\mathop{\rm arg~min}\limits}

\def\bSigma{\mbox{\boldmath $\Sigma$}}

\def\by{\mbox{\boldmath $y$}}
\def\bx{\mbox{\boldmath $x$}}

\def\bp{\mbox{\boldmath $p$}}
\def\bR{\mbox{\boldmath $R$}}

\def\singlebond{\@makechembond\@ne}
\def\doublebond{\@makechembond\tw@}
\def\triplebond{\@makechembond\thr@@}

\lefthead{Umetsu \& Diemer}
\righthead{Mass Profile Shape and Splashback Radius of Galaxy Clusters} 

\usepackage{datetime}
\usepackage{version}    

\usepackage{etoolbox}
\makeatletter
\patchcmd{\NAT@citex}
  {\@citea\NAT@hyper@{\NAT@nmfmt{\NAT@nm}\NAT@date}}
  {\@citea\NAT@nmfmt{\NAT@nm}\NAT@hyper@{\NAT@date}}
  {}
  {}
\patchcmd{\NAT@citex}
  {\@citea\NAT@hyper@{%
     \NAT@nmfmt{\NAT@nm}%
     \hyper@natlinkbreak{\NAT@aysep\NAT@spacechar}{\@citeb\@extra@b@citeb}%
     \NAT@date}}
  {\@citea\NAT@nmfmt{\NAT@nm}%
   \NAT@aysep\NAT@spacechar%
   \NAT@hyper@{\NAT@date}}
  {}
  {}
\patchcmd{\NAT@citex}
  {\@citea\NAT@hyper@{%
     \NAT@nmfmt{\NAT@nm}%
     \hyper@natlinkbreak{\NAT@spacechar\NAT@@open\if*#1*\else#1\NAT@spacechar\fi}%
       {\@citeb\@extra@b@citeb}%
     \NAT@date}}
  {\@citea\NAT@nmfmt{\NAT@nm}%
   \NAT@spacechar\NAT@@open\if*#1*\else#1\NAT@spacechar\fi%
   \NAT@hyper@{\NAT@date}}
  {}
  {}
\makeatother

\begin{document}
\title{Lensing Constraints on the Mass Profile Shape and the Splashback Radius of Galaxy Clusters
\altaffilmark{*}}
\author{Keiichi Umetsu\altaffilmark{1}}   
\author{Benedikt Diemer\altaffilmark{2}}

\altaffiltext{*}
 {Based in part on data collected at the Subaru Telescope,
  which is operated by the National Astronomical Society of Japan.}
\email{keiichi@asiaa.sinica.edu.tw}
\altaffiltext{1}
 {Institute of Astronomy and Astrophysics, Academia Sinica,
  P.O. Box 23-141, Taipei 10617, Taiwan}
\altaffiltext{2}
 {Institute for Theory and Computation, Harvard-Smithsonian Center for
 Astrophysics, 60 Garden St., Cambridge, MA 02138, USA}

\begin{abstract}
The lensing signal around galaxy clusters can, in principle, be
 used to test detailed predictions for their average mass profile from
 numerical simulations. However, the intrinsic shape of the profiles can
 be smeared out when a sample that spans a wide range of cluster masses
 is averaged in physical length units.
 This effect especially conceals rapid changes in gradient such as the
 steep drop associated with the splashback radius, a sharp edge
 corresponding to the outermost caustic in accreting halos.  
 We optimize the extraction of such local features by scaling individual
 halo profiles to a number of spherical overdensity radii, and apply
 this method to 16 X-ray-selected high-mass clusters targeted in the
 Cluster Lensing And Supernova survey with Hubble.
 By forward-modeling the weak- and strong-lensing data presented by
 Umetsu et al., we show that, regardless of the scaling overdensity, 
 the projected ensemble density profile is remarkably well 
 described by a Navarro--Frenk--White (NFW) or Einasto profile out to
 $R\sim 2.5\Mpch$, beyond which the profiles flatten.
 We constrain the NFW concentration to
 $c_\mathrm{200c}=3.66\pm 0.11$ at
 $M_\mathrm{200c}\simeq 1.0\times 10^{15}\Msunh$,
 consistent with and improved from previous work that used
 conventionally stacked lensing profiles, and in 
 excellent agreement with theoretical expectations.
 Assuming the profile form of Diemer \& Kravtsov and generic priors
 calibrated from numerical simulations, we place a lower limit on the
 splashback radius of the cluster halos, if it exists, of
 $\Rsp/r_\mathrm{200m}>0.89$ ($\Rsp>1.83\Mpch$)
 at 68\% confidence. 
 The corresponding density feature is most pronounced when the cluster
 profiles are scaled by $r_\mathrm{200m}$, and smeared out when scaled
 to higher overdensities.  
\end{abstract}   
 
\keywords{cosmology: observations --- dark matter --- galaxies:
clusters: general --- gravitational lensing: strong --- gravitational
lensing: weak}


\section{Introduction}
\label{sec:intro}

Galaxy clusters, as the largest gravitationally bound objects formed in 
the universe, play a fundamental role in our understanding of cosmology
and structure formation. A key ingredient for cluster-based cosmology is
the distribution of dark matter (DM) in and around cluster halos. In
this context, the standard $\Lambda$ cold dark matter ($\Lambda$CDM)
model and its variants, such as self-interacting DM
\citep[SIDM,][]{Spergel+Steinhardt2000} and wave DM
\citep[$\psi$DM,][]{Schive2014psiDM}, 
provide distinct, observationally testable predictions. 

For the case of collisionless DM, high-resolution $N$-body simulations exhibit an
approximately ``universal'' form for the spherically averaged density 
profile of halos in gravitational quasi-equilibrium
\citep[][hereafter NFW]{1997ApJ...490..493N}, 
$\rho(r) \propto (r/r_\mathrm{s})^{-1}(1+r/r_\mathrm{s})^{-2}$,
where $r_\mathrm{s}$ is the characteristic scale radius at which the
logarithmic density slope $d\ln{\rho}/d\ln{r}$ equals $-2$. 
In this context, the halo concentration,
$c_\mathrm{vir}\equiv \rvir/r_\mathrm{s}$,
is a key quantity that characterizes the structure of a halo (all
relevant symbols are defined in detail at the end of this section). 
In the hierarchical $\Lambda$CDM picture of structure formation,
concentration is predicted to correlate with halo mass, 
$\Mvir$, because $r_\mathrm{s}$ stays nearly constant
after an early phase of rapid accretion, whereas $\rvir$ continues to
grow through a mixture of physical accretion and pseudo-evolution
\citep{bullock01profiles, wechsler02haloassembly, cuesta08infall,
zhao09mah, diemer13pe}. As cluster halos are, on average, still
actively growing today, they are expected to have relatively low concentrations, 
$c_\mathrm{vir}\sim 4$--$5$
\citep{Bhatt+2013,Dutton+Maccio2014,Diemer+Kravtsov2015}. These general
trends are complicated by the large scatter in halo growth histories,
which translates into a significant diversity in their density profiles
\citep{Ludlow+2013}. 

Recently, closer examination of the outer halo density profiles in collisionless
$\Lambda$CDM simulations has revealed systematic deviations from the
universal  NFW or \citet{Einasto1965} form
\citep[][hereafter DK14]{Diemer+Kravtsov2014}.
In particular, the profiles exhibit a sharp drop in density,
a feature associated with the last shell that has reached the
apocenter of its first orbit after accreting onto a halo
in spherical collapse models
\citep{gunn72sphericalcollapse, fillmore84, bertschinger85}.
The location of this ``splashback radius,'' $\Rsp$, is within a factor
of two of $r_\mathrm{200m}$ and depends on the mass accretion rate of halos, with a
secondary dependence on redshift \citep[][]{Diemer+Kravtsov2014,
Adhikari2014, More2015splash, Shi2016, More2016splash, adhikari16, mansfield17}. The splashback
radius constitutes a physically motivated halo boundary because (at least in the
spherical case) material outside $\Rsp$ is on its first infall into the
halo, whereas material inside of it is orbiting in the halo potential.

\citet{More2016splash} first observed the splashback feature 
in stacked galaxy surface density profiles around clusters 
(see also \citealt{tully15}; \citealt{patej16}).
Their measured $\Rsp$ is somewhat smaller than expected from the
numerical calibration of \citet{More2015splash}. This intriguing  
disagreement could be due to subtle effects in the analysis,
errors in the numerical calculation, 
baryonic physics affecting cluster member galaxies, or hitherto 
undetected properties of the DM itself, such as self-interaction.

Given this wide range of possible reasons for the disagreement, other
observational probes are of great interest. In particular, we are
looking for a test that is subject to different systematic uncertainties
than cluster member density profiles, but is still applicable to
high-mass galaxy clusters where the splashback signal is expected to be
strongest \citep[][]{Diemer+Kravtsov2014,Adhikari2014}. One potential
probe that fulfills these requirements is gravitational lensing, because 
it measures the total mass profile rather than the distribution of 
subhalos, while the signal is strongest in galaxy clusters.  

Cluster gravitational lensing offers a well-established method for
testing halo structure, through observations of weak shear lensing  
\citep[e.g.,][]{WtG1,WtG3,Gruen2014,Umetsu2014clash,Hoekstra2015CCCP,Okabe+Smith2016,Melchior2016des},
weak magnification lensing
\citep[e.g.,][]{Hildebrandt+2011,Umetsu+2011,Coupon+2013,Ford2014cfhtlens,Jimeno2015,Chiu2016magbias,Ziparo2016locuss},
strong gravitational lensing \citep[e.g.,][]{2005ApJ...621...53B,Zitrin+2009CL0024,Coe+2010,Jauzac2014,Diego2015a1689},
and the combination of all these effects
\citep[e.g.,][]{BTU+05,UB2008,Umetsu+2011stack,Umetsu+2012,Umetsu2015A1689,Umetsu2016clash,Coe+2012A2261,Medezinski+2013}. Over the last decade, cluster lensing observations
\citep{BTU+05,Okabe+2010WL,Okabe+2013,Umetsu+2011stack,Umetsu2016clash,Newman+2013a}  have
established that the projected total mass distribution within individual
and stacked clusters is well described by a family of density profiles  
predicted for cuspy DM-dominated halos, such as the 
NFW \citep{1997ApJ...490..493N}, Einasto \citep{Einasto1965}, and
DARKexp \citep{Hjorth+2010DARKexp,DARKexp2} models. 
Subsequent systematic studies targeting lensing-unbiased cluster samples 
\citep[e.g.,][]{Okabe+2013,Umetsu2014clash,Umetsu2016clash,Merten2015clash,Du2015}
show that the cluster lensing measurements are also in agreement with the
theoretical $c$--$M$ relation that is calibrated for  
recent $\Lambda$CDM cosmologies with a relatively high normalization
\citep{Bhatt+2013,Dutton+Maccio2014,Meneghetti2014clash,Diemer+Kravtsov2015}.

In principle, any feature in the density profiles predicted by
numerical simulations is directly accessible by lensing observations
of a large sample of galaxy clusters 
\citep[e.g.,][]{Okabe+2010WL,Okabe+2013,Umetsu2014clash,Umetsu2016clash,Miyatake2016bias}. 
In reality, however, two effects make such measurements
difficult. First, projecting the density profile into two dimensions
smooths out features because any given sightline crosses a range of
three-dimensional cluster radii. Second, in order to average lensing
observations of individual clusters, we need to stack their density
profiles using some radial scale. Conventionally, physical length units
are chosen for this rescaling
\citep[e.g.,][]{Okabe+2010WL,Sereno2015s8}. If the cluster sample spans
a wide range of masses, such a stacking procedure is likely to smooth
out the intrinsic density profiles of the clusters, and sharp features
such as the splashback radius in particular. Instead, we wish to rescale
the profiles by a halo radius in units of which the features we are
interested in are universal, i.e. they appear at the same rescaled radius.
Numerical simulations show that this choice of scaling radius is far
from trivial: while the inner profiles ($r\simlt \rvir$) are
most universal with halo radii that scale with the critical density of
the universe (such as $r_\mathrm{200c}$), the outer profiles are most
universal when expressed in units of radii that scale with the mean
cosmic density, such as $r_\mathrm{200m}$
\citep[][]{Diemer+Kravtsov2014,Lau2015}. These predictions have not
hitherto been tested observationally.    

In this paper, we develop new methods for scaling and modeling
stacked cluster lensing profiles, and undertake the first investigation
of the splashback radius based on lensing observations. We use the
data presented in \citet[][hereafter U16]{Umetsu2016clash}, who
performed a joint analysis of strong-lensing, weak-lensing shear and
magnification data sets for 20 high-mass clusters targeted in the
Cluster Lensing And Supernova survey with Hubble
\citep[CLASH,][]{Postman+2012CLASH,Umetsu2014clash,Umetsu2016clash,Zitrin2015clash,Merten2015clash}. Their  
analysis combines constraints from 16-band {\em Hubble Space Telescope}
({\em HST}) observations and wide-field multicolor imaging taken
primarily with Suprime-Cam on the Subaru telescope. Such a joint
analysis of multiple lensing probes allows us not only to improve the
precision of mass reconstructions, but also to calibrate systematic
errors inherent in each probe \citep{Rozo+Schmidt2010,Umetsu2013}. The
large radial range covered by the combination of weak- and strong-lensing
data allows us to explore a range of scaling overdensities, and to
investigate their impact on the stacked ensemble fit. Thanks to the
improved stacking procedure, we derive tighter constraints on halo
concentration than in U16, and put a lower limit on the splashback
radius of the stacked CLASH density profile. 

The paper is organized as follows.
In Section \ref{sec:methodology} we summarize the characteristics
of the CLASH sample and describe the data used in this study.
We then outline our procedure for modeling the
cluster lensing profiles, and test its robustness using synthetic
CLASH weak-lensing data.
In Section \ref{sec:results} we apply our methodology to the CLASH
lensing data and fit them with NFW, Einasto, and DK14 profiles.
We discuss the results in Section \ref{sec:discussion}.
Finally, a summary is given in Section \ref{sec:summary}.    

Throughout this paper, we adopt a spatially flat $\Lambda$CDM cosmology 
with
$\Omega_\mathrm{m}=0.27$,
$\Omega_\Lambda=0.73$,
and a Hubble constant 
$H_0 = 100h$\,km\,s$^{-1}$\,Mpc$^{-1}$
with $h=0.7$.
We denote the mean matter density of the universe as
$\rho_\mathrm{m}$ and the critical density as $\rho_\mathrm{c}$. 
We use the standard notation $M_{\Delta_\mathrm{c}}$ or
$M_{\Delta_\mathrm{m}}$ 
to denote the mass enclosed within a sphere of radius
$r_{\Delta_\mathrm{c}}$ or $r_{\Delta_\mathrm{m}}$,
within which the mean overdensity equals 
$\Delta_\mathrm{c} \times \rho_\mathrm{c}(z)$ 
or
$\Delta_\mathrm{m} \times \rho_\mathrm{m}(z)$
at a particular redshift $z$, such that
$M_{\Delta\mathrm{c}}=(4\pi/3)\Delta_\mathrm{c}\rho_\mathrm{c}(z)r_{\Delta\mathrm{c}}^3$
and
$M_{\Delta\mathrm{m}}=(4\pi/3)\Delta_\mathrm{m}\rho_\mathrm{m}(z)r_{\Delta\mathrm{m}}^3$.
We generally denote three-dimensional cluster radii as $r$, and reserve
the symbol $R$ for projected clustercentric distances. We define the
splashback radius of a three-dimensional density profile, $\Rsp$, as the
radius where the logarithmic slope of the profile is steepest.
Similarly, we use $R_\mathrm{sp}^\mathrm{2D}$ to denote the splashback
radius derived from the steepest slope of the projected profile. 
We compute the virial mass and radius, $\Mvir$ and $\rvir$,
using an expression for $\Delta_\mathrm{vir}(z)$ based on the
spherical collapse model \citep[Appendix A of][]{1998PASJ...50....1K}.
For a given overdensity $\Delta$, the concentration parameter is
defined as $c_\Delta=r_\Delta/r_\mathrm{s}$.
All quoted errors are $1\sigma$ confidence limits (CL) unless otherwise
stated.


\section{Data and Methodology}
\label{sec:methodology}

\subsection{CLASH Sample and Data}
\label{subsec:data}

\begin{deluxetable*}{lccccccccc}
\tablecolumns{10}
\centering
\tablewidth{0pt}
\tabletypesize{\scriptsize}
\tablecaption{
\label{tab:sample}
Properties of the Cluster Sample
}
\tablehead{
 \multicolumn{1}{l}{Cluster} &
 \multicolumn{1}{c}{$z_\mathrm{l}$} &
 \multicolumn{1}{c}{$M_\mathrm{2500c}$} &
 \multicolumn{1}{c}{$c_\mathrm{2500c}$} &
 \multicolumn{1}{c}{$M_\mathrm{500c}$} &
 \multicolumn{1}{c}{$c_\mathrm{500c}$} &
 \multicolumn{1}{c}{$M_\mathrm{vir}$} &
 \multicolumn{1}{c}{$c_\mathrm{vir}$} &
 \multicolumn{1}{c}{$M_\mathrm{200m}$} &
 \multicolumn{1}{c}{$c_\mathrm{200m}$} 
 \\ 
\multicolumn{1}{l}{} &
\multicolumn{1}{c}{} &
\multicolumn{1}{c}{$(10^{14}M_\odot/h)$} &
\multicolumn{1}{c}{} &
\multicolumn{1}{c}{$(10^{14}M_\odot/h)$} &
\multicolumn{1}{c}{} &
\multicolumn{1}{c}{$(10^{14}M_\odot/h)$} &
\multicolumn{1}{c}{} &
\multicolumn{1}{c}{$(10^{14}M_\odot/h)$} &
\multicolumn{1}{c}{} 
}
\startdata
           Abell 383 & $0.187$ & $ 1.95 \pm  0.44$ & $ 1.83 \pm  0.65$ & $ 4.11 \pm  1.21$& $ 3.97 \pm  1.24$ & $ 6.59 \pm  2.33$ & $ 7.63 \pm  2.22$ & $ 7.24 \pm  2.64$ & $ 8.91 \pm  2.55$\\
           Abell 209 & $0.206$ & $ 2.06 \pm  0.48$ & $ 0.68 \pm  0.19$ & $ 6.75 \pm  1.38$& $ 1.71 \pm  0.40$ & $13.72 \pm  3.23$ & $ 3.54 \pm  0.73$ & $15.65 \pm  3.81$ & $ 4.16 \pm  0.84$\\
          Abell 2261 & $0.224$ & $ 4.14 \pm  0.72$ & $ 1.03 \pm  0.31$ & $10.95 \pm  2.13$& $ 2.43 \pm  0.62$ & $19.74 \pm  4.81$ & $ 4.82 \pm  1.11$ & $21.99 \pm  5.56$ & $ 5.60 \pm  1.27$\\
   RX~J2129.7$+$0005 & $0.234$ & $ 1.45 \pm  0.31$ & $ 1.71 \pm  0.59$ & $ 3.14 \pm  0.81$& $ 3.75 \pm  1.12$ & $ 5.05 \pm  1.56$ & $ 7.15 \pm  1.98$ & $ 5.51 \pm  1.75$ & $ 8.21 \pm  2.25$\\
           Abell 611 & $0.288$ & $ 2.89 \pm  0.64$ & $ 1.10 \pm  0.42$ & $ 7.51 \pm  1.86$& $ 2.56 \pm  0.82$ & $13.26 \pm  4.04$ & $ 4.97 \pm  1.45$ & $14.54 \pm  4.58$ & $ 5.64 \pm  1.62$\\
         MS2137$-$2353 & $0.313$ & $ 1.73 \pm  0.50$ & $ 0.68 \pm  0.42$ & $ 5.80 \pm  1.80$& $ 1.71 \pm  0.87$ & $11.88 \pm  5.12$ & $ 3.44 \pm  1.56$ & $13.24 \pm  5.98$ & $ 3.90 \pm  1.74$\\
   RX~J1532.9$+$3021 & $0.348$ & $ 3.11 \pm  0.74$ & $ 0.98 \pm  0.48$ & $ 8.71 \pm  2.54$& $ 2.32 \pm  0.96$ & $15.78 \pm  6.15$ & $ 4.48 \pm  1.69$ & $17.17 \pm  6.94$ & $ 5.01 \pm  1.86$\\
   RX~J2248.7$-$4431 & $0.352$ & $ 2.45 \pm  0.58$ & $ 0.77 \pm  0.26$ & $ 7.47 \pm  1.56$& $ 1.90 \pm  0.53$ & $14.21 \pm  3.48$ & $ 3.75 \pm  0.93$ & $15.57 \pm  3.92$ & $ 4.19 \pm  1.02$\\
 MACS~J1115.9$+$0129 & $0.352$ & $ 2.87 \pm  0.79$ & $ 1.26 \pm  0.69$ & $ 7.36 \pm  2.83$& $ 2.87 \pm  1.36$ & $12.61 \pm  6.34$ & $ 5.44 \pm  2.35$ & $13.61 \pm  7.05$ & $ 6.05 \pm  2.59$\\
 MACS~J1931.8$-$2635 & $0.363$ & $ 1.28 \pm  0.70$ & $ 1.59 \pm  0.99$ & $ 2.92 \pm  1.20$& $ 3.47 \pm  1.94$ & $ 4.93 \pm  1.95$ & $ 6.44 \pm  3.36$ & $ 5.31 \pm  2.14$ & $ 7.13 \pm  3.69$\\
 MACS~J1720.3$+$3536 & $0.391$ & $ 2.75 \pm  0.60$ & $ 1.16 \pm  0.46$ & $ 6.97 \pm  1.77$& $ 2.68 \pm  0.90$ & $11.93 \pm  3.77$ & $ 5.05 \pm  1.55$ & $12.81 \pm  4.16$ & $ 5.58 \pm  1.69$\\
 MACS~J0429.6$-$0253 & $0.399$ & $ 2.00 \pm  0.49$ & $ 1.36 \pm  0.60$ & $ 4.80 \pm  1.47$& $ 3.06 \pm  1.17$ & $ 7.95 \pm  3.03$ & $ 5.70 \pm  2.00$ & $ 8.49 \pm  3.32$ & $ 6.27 \pm  2.18$\\
 MACS~J1206.2$-$0847 & $0.440$ & $ 3.24 \pm  0.71$ & $ 1.03 \pm  0.39$ & $ 8.57 \pm  1.74$& $ 2.43 \pm  0.77$ & $14.94 \pm  3.71$ & $ 4.58 \pm  1.33$ & $15.97 \pm  4.07$ & $ 5.00 \pm  1.43$\\
 MACS~J0329.7$-$0211 & $0.450$ & $ 2.31 \pm  0.45$ & $ 2.12 \pm  0.59$ & $ 4.55 \pm  0.96$& $ 4.53 \pm  1.12$ & $ 6.81 \pm  1.61$ & $ 8.11 \pm  1.89$ & $ 7.15 \pm  1.72$ & $ 8.79 \pm  2.03$\\
   RX~J1347.5$-$1145 & $0.451$ & $ 5.35 \pm  1.14$ & $ 0.85 \pm  0.31$ & $15.63 \pm  3.42$& $ 2.07 \pm  0.62$ & $28.46 \pm  7.79$ & $ 3.95 \pm  1.06$ & $30.51 \pm  8.57$ & $ 4.31 \pm  1.15$\\
 MACS~J0744.9$+$3927 & $0.686$ & $ 3.02 \pm  0.72$ & $ 0.95 \pm  0.40$ & $ 8.36 \pm  1.97$& $ 2.27 \pm  0.81$ & $14.40 \pm  4.19$ & $ 4.11 \pm  1.34$ & $14.93 \pm  4.40$ & $ 4.30 \pm  1.39$\\
\hline
    Stacked ensemble & $0.337$ & $ 2.64 \pm  0.14$ & $ 1.03 \pm  0.10$ & $ 6.87 \pm  0.43$& $ 2.43 \pm  0.20$ & $11.99 \pm  0.93$ & $ 4.69 \pm  0.35$ & $13.03 \pm  1.04$ & $ 5.26 \pm  0.38$
\enddata
\tablecomments{
All mass estimates listed are based on the combined strong-lensing,
 weak-lensing shear and magnification analysis of 
 \citet{Umetsu2016clash}. See Table 2 of \citet{Umetsu2016clash} for
 $M_\mathrm{200c}$ and $c_\mathrm{200c}$. The effective $M_\Delta$
 masses of the sample were extracted from a spherical NFW fit to their
 stacked surface mass density profile \citep{Umetsu2016clash}.  
 The best-fit model yields
 $M_\mathrm{200c}=(10.11 \pm 0.73)\times 10^{14}M_\odot\,h^{-1}$ and 
 $c_\mathrm{200c}=3.76\pm 0.28$ 
 \citep[symmetrized errors; see the NFW model in Table 4 of][]{Umetsu2016clash}.
 The effective redshift of the sample represents a sensitivity-weighted
 average of their redshifts 
 \citep[Section 5.4 of][]{Umetsu2016clash}, $0.337$,
 which is close to the median redshift, $\overline{z}_\mathrm{l}=0.352$.
} 
\end{deluxetable*}

The CLASH survey \citep{Postman+2012CLASH} is a 524-orbit {\em HST}
Multi-Cycle  Treasury program targeting 25 high-mass galaxy clusters.
Of these, 20 CLASH clusters were selected to be X-ray hot ($T_X>5$\,keV) 
and to have a high degree of regularity in their X-ray morphology,
with no lensing information used a priori. 
Another subset of five clusters were selected by their
high-magnification properties. These high-magnification clusters often
turn out to be complex, massive merging systems
\citep[e.g.,][]{Zitrin+2013M0416,Medezinski+2013}. 
A complete definition of the CLASH sample is given in
\citet{Postman+2012CLASH}. 

In this work, we shall focus on the analysis of the X-ray-selected
subsample to simplify the interpretation of our results. Numerical
simulations suggest that this subsample is mostly composed of relaxed
clusters ($\sim 70\%$) and largely free of orientation bias
\citep{Meneghetti2014clash}. Specifically, we use a lensing-unbiased
subset of 16 CLASH X-ray-selected clusters taken from
U16, who performed a comprehensive analysis of the
strong-lensing, weak-lensing shear and magnification data.
Our cluster sample lies in the redshift range 
$0.19\simlt z_\mathrm{l}\simlt 0.69$ and over a mass range
$5\simlt M_\mathrm{vir}/(10^{14}\Msunh)\simlt
30$
(Table \ref{tab:sample}),
spanning a factor of $\sim 6$ in halo mass $M_\Delta$, or a factor of
$\sim 1.8$ in $r_\Delta\propto M_\Delta^{1/3}$.
A full description of the data used in this work is given by
U16. Here, we provide only a brief summary of the
most relevant aspects of the lensing reconstructions. 

The U16 analysis
uses the cluster lensing mass inversion (\textsc{clumi}) code developed
by \citet{Umetsu+2011} and \citet{Umetsu2013}, in which lensing
constraints are combined a posteriori in the form of azimuthally
averaged radial profiles. U16 used constraints spanning the radial range 
$10\arcsec$--$960\arcsec$ obtained from 16-band {\em HST} observations
\citep{Zitrin2015clash} and wide-field multicolor imaging
\citep{Umetsu2014clash} taken primarily with Subaru/Suprime-Cam.
The position of the brightest cluster galaxy (BCG) is adopted as the
cluster center (Table 1 in U16). 
\citet{Umetsu2014clash} obtained weak-lensing shear and magnification
measurements in 10 logarithmically spaced radial bins ($N_\mathrm{WL}=10$) over the
range $0.9\arcmin$--$16\arcmin$ for all clusters observed with Subaru,
and $0.9\arcmin$--$14\arcmin$ for RX~J2248.7$-$4431 observed with ESO/WFI.
\citet{Zitrin2015clash} constructed detailed mass models for each
cluster core from a joint analysis of {\em HST} strong and weak-shear
lensing data.
U16 constructed enclosed projected mass constraints
for a set of four equally-spaced integration radii 
($10\arcsec$--$40\arcsec$, $N_\mathrm{SL}=4$)
from the {\em HST} lensing analysis of \citet{Zitrin2015clash}.
U16 combined these full lensing constraints for
individual clusters in their joint likelihood analysis to reconstruct
binned surface mass density profiles
$\bSigma=\{\Sigma(R_i)\}_{i=1}^{N_\mathrm{bin}}$ 
measured in a set of clustercentric radial bins,
$\bR=\{R_i\}_{i=1}^{N_\mathrm{bin}}$ with
$N_\mathrm{bin}=N_\mathrm{SL}+N_\mathrm{WL}+1$. We have
$N_\mathrm{bin}=15$ bins for all clusters,
except $N_\mathrm{bin}=11$ for RX\,J1532.9$+$3021 with $N_\mathrm{SL}=0$,
for which no secure identification of multiple images was made
\citep{Zitrin2015clash}. The $\bSigma$ profiles used in this work are
shown in Figure 11 of U16.

U16 accounted for various sources of errors.
Their analysis includes four terms in the total covariance matrix
$C_{ij}$ of the $\bSigma$ profile,  
\begin{equation}
 \label{eq:Ctot}
 C = C^\mathrm{stat} + C^\mathrm{sys} + C^\mathrm{lss} + C^\mathrm{int},
\end{equation}
where 
$C^\mathrm{stat}$ represents statistical observational errors,
$C^\mathrm{sys}$ contains systematic uncertainties due primarily to
the residual mass-sheet uncertainty \citep{Umetsu2014clash},
$C^\mathrm{lss}$ is the cosmic-noise covariance matrix due to
projected uncorrelated large-scale structures
\citep{2003MNRAS.339.1155H,Umetsu+2011stack}, 
and 
$C^\mathrm{int}$ accounts for the intrinsic variations of the projected 
cluster lensing signal at fixed mass due to variations in halo
concentration, cluster asphericity, and the presence of correlated halos
\citep{Gruen2015}. 
Overall, the reconstruction uncertainty is dominated by the
$C^\mathrm{stat}$ term (Figure 1 of U16).
The relative contribution from the $C^\mathrm{int}$ term becomes
increasingly dominant at small cluster radii, especially at
$\theta\simlt 2\arcmin$. 
The impact of the $C^\mathrm{lss}$ term is most important at large
cluster radii, where the cluster signal is small. 

Table \ref{tab:sample} summarizes CLASH lensing determinations of the
mass and concentration parameters $(M_\Delta,c_\Delta)$
for our 16 clusters based on the full lensing
analysis of U16.
These values were obtained from spherical NFW fits to individual cluster 
$\bSigma$ profiles using the total covariance matrix $C$ (Equation
(\ref{eq:Ctot})), 
restricting the fitting range to $R\le 2\Mpch$
($\sim 2r_\mathrm{500c}\sim r_\mathrm{200m}$
for the CLASH sample) to avoid systematic effects
\citep{Becker+Kravtsov2011,Meneghetti2014clash}. 
In Table \ref{tab:sample}, we also list effective overdensity masses 
$M_\Delta^\mathrm{eff}$ of the sample,
which were obtained by U16 from a spherical NFW fit
to the stacked surface mass density profile of the 16 CLASH clusters.
The stacked ensemble has an effective halo mass
$M_\mathrm{vir}^\mathrm{eff}=(11.99\pm 0.93)\times 10^{14}\Msunh$ and
an effective halo concentration
$c_\mathrm{vir}^\mathrm{eff}=4.69\pm 0.35$
($c_\mathrm{200c}^\mathrm{eff}=3.76\pm 0.28$; see Table \ref{tab:sample}),
and lies at a sensitivity-weighted average redshift of
$z_\mathrm{l}^\mathrm{eff}=0.337$, close to the median
redshift of $\overline{z}_\mathrm{l} = 0.352$.

U16 quantified potential sources of 
systematic uncertainty in their mass calibration (see their Section 7.1), 
such as the effect of dilution of the weak-lensing signal by cluster
members (2.4\%), photometric-redshift bias (0.27\%), 
shear calibration uncertainty (5\%),
and
projection effects of prolate halos (3\%).
Combining them in quadrature, the total systematic uncertainty in the 
absolute mass calibration was estimated to be $\simeq 6\%$.
This is in close agreement with the value $\sim 8\%$ empirically
estimated from the shear-magnification consistency test of
\citet{Umetsu2014clash}. 

Another potential source of systematic errors is smoothing of the
central lensing signal from miscentering effects 
\citep{Johnston+2007a,Umetsu+2011stack,Du+Fan2014}.
On average, the sample exhibits a small positional offset between the
BCG and the X-ray peak, characterized by an rms offset of
$\sigma_\mathrm{off}\simeq 11\kpch$ \citep[][]{Umetsu2014clash,Umetsu2016clash}, 
which is much smaller than the typical resolution limit of our {\em HST}
lensing data ($\theta_\mathrm{min}=10\arcsec$, corresponding to
$\simeq 35\kpch$ at $\overline{z}_\mathrm{l}=0.35$).
Hence, the miscentering effects are not expected to significantly affect 
our ensemble lensing analysis.

\subsection{Radial Scaling of the Profiles}
\label{subsec:scaling}

One of our main goals in this work is to investigate how the radial
scaling of stacked profiles influences the fit results. Ideally, one
would like to stack profiles as a function of the exact halo radius
where a particular feature is expected, e.g. the NFW scale radius or the
splashback radius. However, since their locations are {\em a priori}
unknown, 
we need to resort to an alternative halo radius that can be measured for
individual clusters, generally a spherical overdensity radius
$r_\Delta$. Now, the goal is to choose a definition in which the
location of the feature in question is {\em universal}, i.e. independent
of halo mass, and possibly of redshift. 

Unfortunately, there is no guarantee that any one definition will be
ideal for multiple features. In fact, DK14 discovered that halo
density profiles in $N$-body simulations prefer different scaling radii
in different regions of the profile: while the inner profiles (and thus
concentrations) are most universal when expressed in units of halo radii 
that scale with the critical density $\rho_\mathrm{c}(z)$ of the 
universe, such as $r_\mathrm{200c}$, the outer profiles (and thus the
splashback radius) are most universal in units of halo radii that scale
with the mean cosmic density $\rho_\mathrm{m}(z)$, such as
$r_\mathrm{200m}$. These scalings were confirmed in hydrodynamical
simulations of galaxy clusters \citep[][see also \citealt{Shi2016}]{Lau2015}. As we wish to investigate
both the inner and outer profiles, we repeat our analysis using a number
of different scalings covering a range between $2500\rho_\mathrm{c}$ and
$200\rho_\mathrm{m}\simeq 94\rho_\mathrm{c}$ at 
$z_\mathrm{l}^\mathrm{eff}= 0.337$.

To construct the scaled surface mass density profiles for the CLASH
sample, we use our full lensing constraints on the NFW 
parameters $M_\Delta$ and $c_\Delta$ of each individual cluster as obtained by U16 (Section
\ref{subsec:data}; see Table \ref{tab:sample}), 
and normalize their observed $\Sigma(R)$ profiles to a given overdensity
$\Delta$ of interest.
We stress that we do not rely on scaling relations 
(e.g., the $c$--$M$ relation), but use observational lensing constraints 
on $r_\Delta$ and $\Sigma(r_\Delta)$ for each cluster.
The high-quality, multiscale weak- and strong-lensing
constraints from the CLASH survey enable us to explore the wide range of 
overdensities listed above.

We choose the NFW model for the scaling because recent cluster lensing
observations show that the ``projected total'' matter distribution in
the intracluster region ($R\simlt r_\mathrm{200m}$) is in excellent
agreement with the NFW form 
\citep[][]{Umetsu+2011stack,Umetsu2014clash,Umetsu2016clash,Silva+2013,Newman+2013a,Okabe+2013,Niikura2015,Okabe+Smith2016},
as predicted for collisionless DM-dominated halos in $N$-body
cosmological simulations \citep{Oguri+Hamana2011,Meneghetti2014clash}. We
demonstrate the consistency of this choice with our results in Section \ref{sec:results}.

\subsection{Fitting Functions}
\label{subsec:DK14}

The second goal of this work (besides investigating radial
scalings) is to constrain the splashback radius of the CLASH cluster
sample. While NFW and Einasto fitting functions were sufficient for the
radial rescaling, they describe only the 1-halo term and do not take the
steepening due to the splashback radius into account. Thus, we also fit
the cluster lensing profiles with the more flexible fitting function of
DK14 which was calibrated to a suite of $\Lambda$CDM $N$-body
simulations.
We emphasize that the quality of the data is insufficient to distinguish
among the different profile models, which all describe the data very
well (see Section \ref{subsec:fit_quality}).
Nevertheless, in order to constrain the location of the splashback
radius, we {\em assume} that the $\Lambda$CDM simulations of DK14
describe real cluster halos and use the DK14 profile as a
fitting function in conjunction with generic priors.

Furthermore, we note that the ``true'' location of the splashback radius
is not strictly equivalent to a particular location in the spherically
averaged density profile. However, we follow \citet{More2015splash} in
defining $\Rsp$ as the radius where the logarithmic slope of the
three-dimensional density profile is steepest. According to this
definition, $\Rsp$ is expected to lie within a factor of two of
$r_\mathrm{200m}$ \citep{More2015splash}. 
Furthermore, the steepest slope would need to be steeper than that
expected from the sum of the Einasto profile and the 2-halo term at 
large scales if a detection were to be claimed \citep{More2016splash}.

The DK14 fitting formula is described by eight parameters, and is
sufficiently flexible to reproduce a range of fitting functions for the
DM density profile, such as the halo model
\citep{Oguri+Hamana2011,Hikage2013}. We use the publicly available code  
\textsc{colossus} 
\citep{colossus} for many of calculations relating to density
profiles. The DK14 model is given by  
\begin{equation}
 \label{eq:DK14}
 \begin{aligned}
\Delta\rho(r) &= \rho(r)-\rho_\mathrm{m} =\rho_\mathrm{inner}\times f_\mathrm{trans}+\rho_\mathrm{outer},\\
\rho_\mathrm{inner} &= \rho_\mathrm{Einasto} = \rho_\mathrm{s}
 \exp\left\{-\frac{2}{\alpha}\left[
\left(\frac{r}{r_\mathrm{s}}\right)^\alpha-1\right]\right\},\\
f_\mathrm{trans} &= \left[1+
		      \left(\frac{r}{r_\mathrm{t}}\right)^\beta\right]^{-\frac{\gamma}{\beta}},\\
\rho_\mathrm{outer}&=\frac{b_\mathrm{e}\rho_\mathrm{m}}{\Delta_\mathrm{max}^{-1}
  + (r/r_\mathrm{piv})^{s_\mathrm{e}}},\\
 \end{aligned}
\end{equation}
with
$r_\mathrm{piv}=5r_\mathrm{200m}$ and
$\Delta_\mathrm{max} = 10^3$. 
Here $\Delta_\mathrm{max}$ has been introduced as a maximum cutoff
density of the outer term to avoid a spurious contribution at small halo
radii \citep{colossus}. 
The Einasto profile $\rho_\mathrm{Einasto}$ describes the intracluster
mass distribution, with the shape parameter $\alpha$ describing the
degree of profile curvature and $r_\mathrm{s}$ the scale radius at which
the logarithmic slope is -2. 
The transition term $f_\mathrm{trans}$ characterizes the steepening 
around a truncation radius, $r_\mathrm{t}$.  
The outer term $\rho_\mathrm{outer}$, given by a softened power law, is
responsible for the correlated matter distribution around clusters, also
known as the 2-halo term, $\rho_\mathrm{2h}$. 
DK14 found that this fitting function provides a precise description
($\simlt 5\%$) of their simulated DM density profiles at
$r\simlt 9r_\mathrm{vir}$.
At larger radii, the outer term is expected to follow a shape
proportional to the matter correlation function
\citep[e.g.,][]{Oguri+Takada2011}.

To scale out the mass dependence of the density profile, we use an arbitrary
overdensity radius $r_\Delta$ as a pivot radius, and define
a scaled version of $\Delta\rho(r)$ as a function of $r=r_\Delta x$, namely
\begin{equation} 
 \begin{aligned}
  \Delta\rho(r)
  \propto&
  \exp\left[ -\frac{2}{\alpha}c_\Delta^\alpha(x^\alpha-1) \right] 
  \left[1+ \left(\frac{x}{\tau_\Delta}\right)^\beta\right]^{-\frac{\gamma}{\beta}}\\
  &+ \frac{B_\Delta}{\epsilon_{\Delta} + x^{s_\mathrm{e}}} \equiv f_{\Delta}(x),
 \end{aligned}
\end{equation}
with $\epsilon_{\Delta}=\Delta_\mathrm{max}^{-1}\times(5r_\mathrm{200m}/r_\Delta)^{s_\mathrm{e}}$.\footnote{For $\Delta=\mathrm{200m}$, $\epsilon_\Delta$ is constant for all clusters,
$\epsilon_\mathrm{200m}\simeq0.11$ at $s_\mathrm{e}=1.5$; otherwise, the
actual value of $\epsilon_\Delta$ depends on the ratio 
$r_\mathrm{200m}/r_\Delta$ of each individual cluster.}
This scaled DK14 model is described by a set of seven dimensionless parameters,
\begin{equation}
 \bp=\{c_\Delta,\alpha,\tau_\Delta,B_\Delta,s_\mathrm{e},\beta,\gamma\},
\end{equation}
namely the halo concentration $c_\Delta=r_\Delta/r_\mathrm{s}$,
the Einasto shape parameter $\alpha$,
the dimensionless truncation radius
$\tau_\mathrm{\Delta}=r_\mathrm{t}/r_\mathrm{\Delta}$,
the relative normalization of the outer term $B_\Delta\propto b_\mathrm{e}$,
and three additional shape parameters ($s_\mathrm{e},\beta,\gamma$) for
the transition and outer terms. The model reduces to the Einasto model
(specified by $c_\Delta$ and $\alpha$)
when $f_\mathrm{trans}=1$ ($\tau_\Delta\to \infty$) and
$f_\mathrm{outer}=0$ ($B_\Delta=0$). We refer the reader to Appendix \ref{appendix:dk14}
for a more detailed description of the scaled DK14 profile function. 
Similarly, we define a scaled version of the NFW profile, 
$\Delta\rho(r)\propto [c_\Delta x (1+c_\Delta x)^2]^{-1}$ with $r=r_\Delta x$, 
described by the concentration parameter $c_\Delta$.

By projecting $\Delta\rho(r)$ along the line of sight, we derive the
scaled surface mass density, which is a lensing observable,
\begin{equation}
 \label{eq:ydef}
 y_\Delta(x) := \frac{\Sigma(R=r_\mathrm{\Delta}x)}{\Sigma(r_\mathrm{\Delta})},
\end{equation}
normalized as $y_\Delta(x)=1$ at $x=1$, where 
\begin{equation}
 \Sigma(R) 
=  2\int_{R}^\infty \frac{\Delta\rho(r)rdr}{\sqrt{r^2-R^2}}.
\end{equation}
The projected $y_\Delta(x)$ profile is modeled in terms of the scaled
density profile $f_{\Delta}$ as
\begin{equation}
 \begin{aligned}
  y_\Delta(x|\bp) = \left(\int_x^\infty\!\frac{f_{\Delta}(\xi|\bp)\xi
  d\xi}{\sqrt{\xi^2-x^2}}\right)
   \left(\int_1^\infty\!\frac{f_{\Delta}(\xi|\bp)\xi
  d\xi}{\sqrt{\xi^2-1}}\right)^{-1}.
 \end{aligned}
\end{equation}
We note that it is straightforward to generalize our approach to
shear-only weak-lensing observations where the {\em differential}
surface mass density $\Delta\Sigma(R)=\Sigma(<R)-\Sigma(R)$ is a direct
observable in the weak-lensing limit \citep[e.g.,][]{2001PhR...340..291B}.

\subsection{Parameter Inference}
\label{subsec:bayesian}

We use a Bayesian approach to infer the shape and structural parameters 
of the mass distribution of the CLASH sample.
We restrict the optimization to ($c_\Delta, \alpha, \tau_\Delta, B_\Delta$),
the primary model parameters that describe the scaled DK14 model, and 
marginalize over the remaining parameters, $(s_\mathrm{e},\beta,\gamma)$,
using priors based on the $N$-body simulations of DK14.
In particular, following \citet{More2016splash}, we adopt 
Gaussian priors of
$\log_{10}\beta=\log_{10}{6}\pm 0.2$ and 
$\log_{10}\gamma=\log_{10}{4}\pm 0.2$\footnote{This corrects
typographical errors in Section 2 of \citet{More2016splash}. 
See also Section 3.1 of \citet{More2015splash}.},
allowing a wide, representative range of values. 
We assume a Gaussian prior on $s_\mathrm{e}$ of $1.5\pm 0.1$, centered 
around the value found by DK14.
We use unconstraining flat priors on $(c_\Delta, \alpha, B_\Delta)$.
For $\tau_\Delta$, we assume $\tau_\mathrm{200m}\in [0,5]$, where the 
upper bound corresponds approximately to 
$r_\mathrm{t}=5r_\mathrm{200m}\simeq 10\Mpch$ for the CLASH 
sample (Table \ref{tab:sample}),
which is larger than the maximum data radius of
$\sim 5\Mpch$.
We translate this prior to a given overdensity $\Delta$ using the
effective $r_\Delta$ radius of the sample (Section \ref{subsec:data} and
Table \ref{tab:sample}) as
$\tau_\Delta\in[0, 5 (r_\mathrm{200m}^\mathrm{eff}/r_\Delta^\mathrm{eff})]$.

We use a Gaussian log-likelihood 
$-\ln{\cal L}(\bp)=\chi^2(\bp)/2+\mathrm{const.}$ with a $\chi^2$ 
function given by
\begin{equation}
 \begin{aligned}
  \chi^2 = \sum_{n=1}^{N_\mathrm{halo}}\sum_{i,j=1}^{N_\mathrm{bin}}&
  \left[y_{\Delta,i}^{(n)}-y_\Delta(x^{(n)}_i|\bp)\right]
    \left[C^{(n)}_{\Delta}\right]^{-1}_{ij}\\
  \times& \left[y_{\Delta,j}^{(n)}-y_\Delta(x^{(n)}_j|\bp)\right],
 \end{aligned}
\end{equation}
where
$\by_\Delta=\{y_{\Delta,i}\}_{i=1}^{N_\mathrm{bin}}$
is the scaled data vector (see Equation (\ref{eq:ydef})) for each cluster
sampled at $\bx=\{x_i\}_{i=1}^{N_\mathrm{bin}}=\{R_i/r_\Delta\}_{i=1}^{N_\mathrm{bin}}$,
$C_{\Delta}$ is the total covariance matrix of the scaled data
$\by_\Delta$,  
and $y_\Delta(x_i|\bp)$ represents the model for $y_{\Delta,i}$
with parameters $\bp$.
Combining all 16 clusters in our sample, we have a total of $236$ data
points (Section \ref{subsec:data}).

We use a Markov chain Monte Carlo (MCMC) approach with
Metropolis--Hastings sampling to sample from the posterior distribution
of the parameters,   
$(c_\Delta,\alpha,\tau_\Delta,B_\Delta,s_\mathrm{e},\log_{10}\beta,\log_{10}\gamma)$,
given the data and the priors stated above.
We largely follow the sampling procedure of \citet{Dunkley+2005} but
employ the Gelman--Rubin statistic \citep{Gelman+Rubin1992}
as a convergence criterion of the generated chains. Once convergence to
a stationary distribution is achieved, we run a long, final chain of
$10^5$ sampled points, which is used for our parameter estimation and
error analysis. 
We also use the \textsc{minuit} package from the CERN program libraries
to find the global maximum a posteriori estimate of the joint
probability distribution. This procedure allows a further refinement of
the {\em best-fit} solution with respect to the one obtained from the
MCMC sampling \citep[see discussions
in][]{Planck2014XVI,Penna-Lima2016}.

\subsection{Tests with Synthetic Data}
\label{subsec:mock}


\begin{figure*}[!htb]
 \begin{center}
  $ 
  \begin{array}
   {c@{\hspace{.3in}}c}
    \includegraphics[scale=0.45, angle=0,clip]{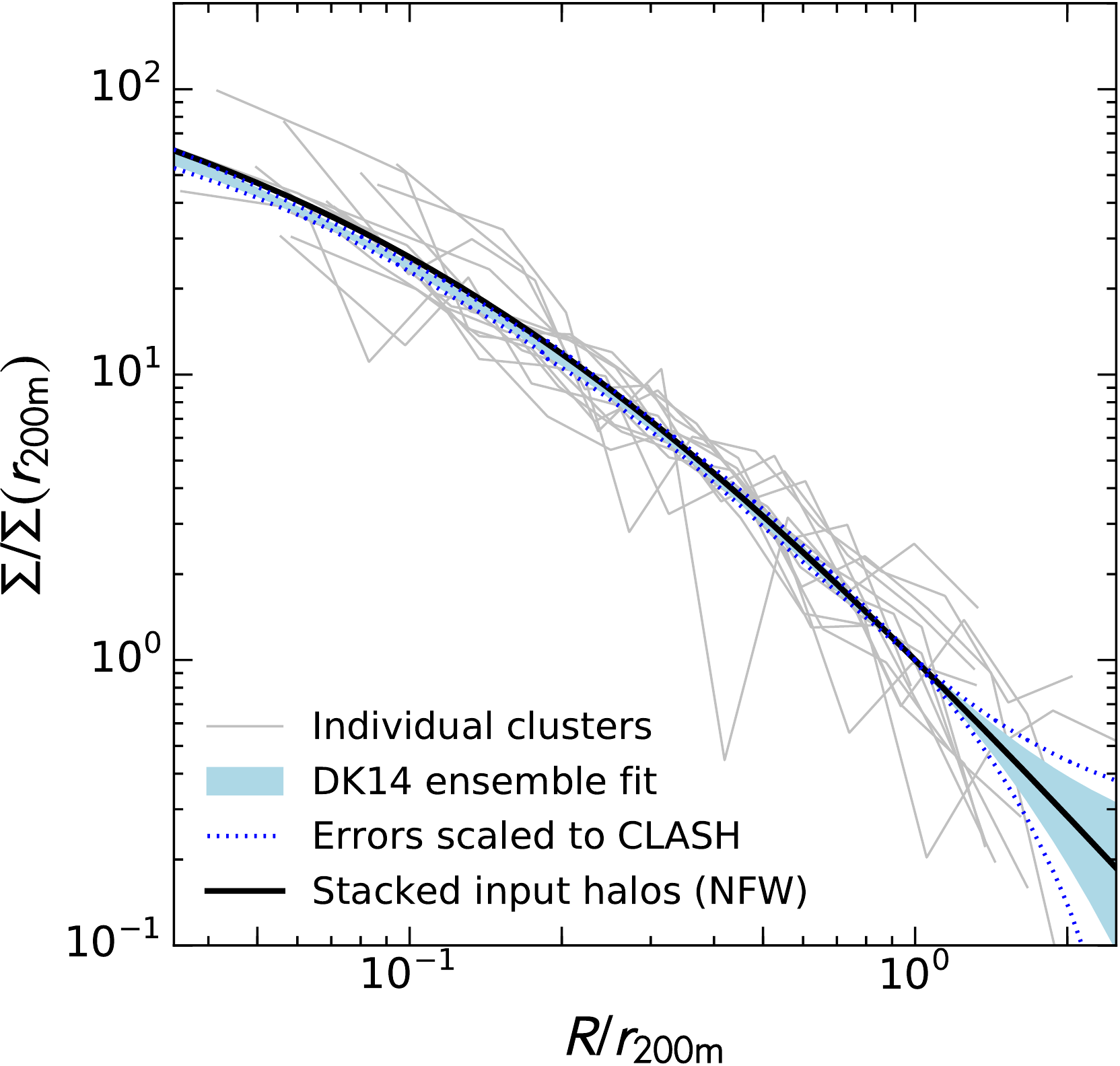} & 
    \includegraphics[scale=0.45, angle=0,clip]{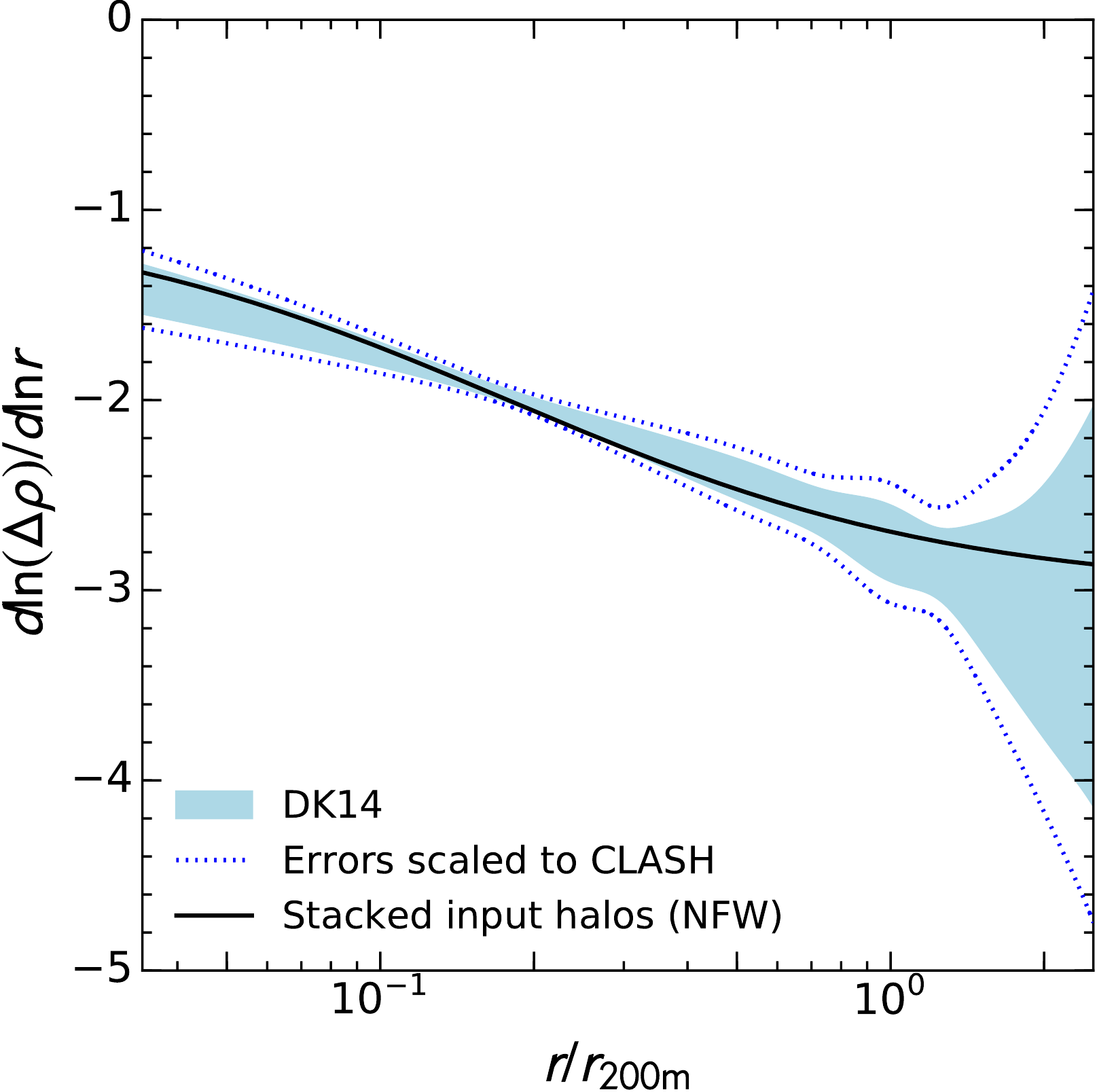}\\
    \includegraphics[scale=0.45, angle=0,clip]{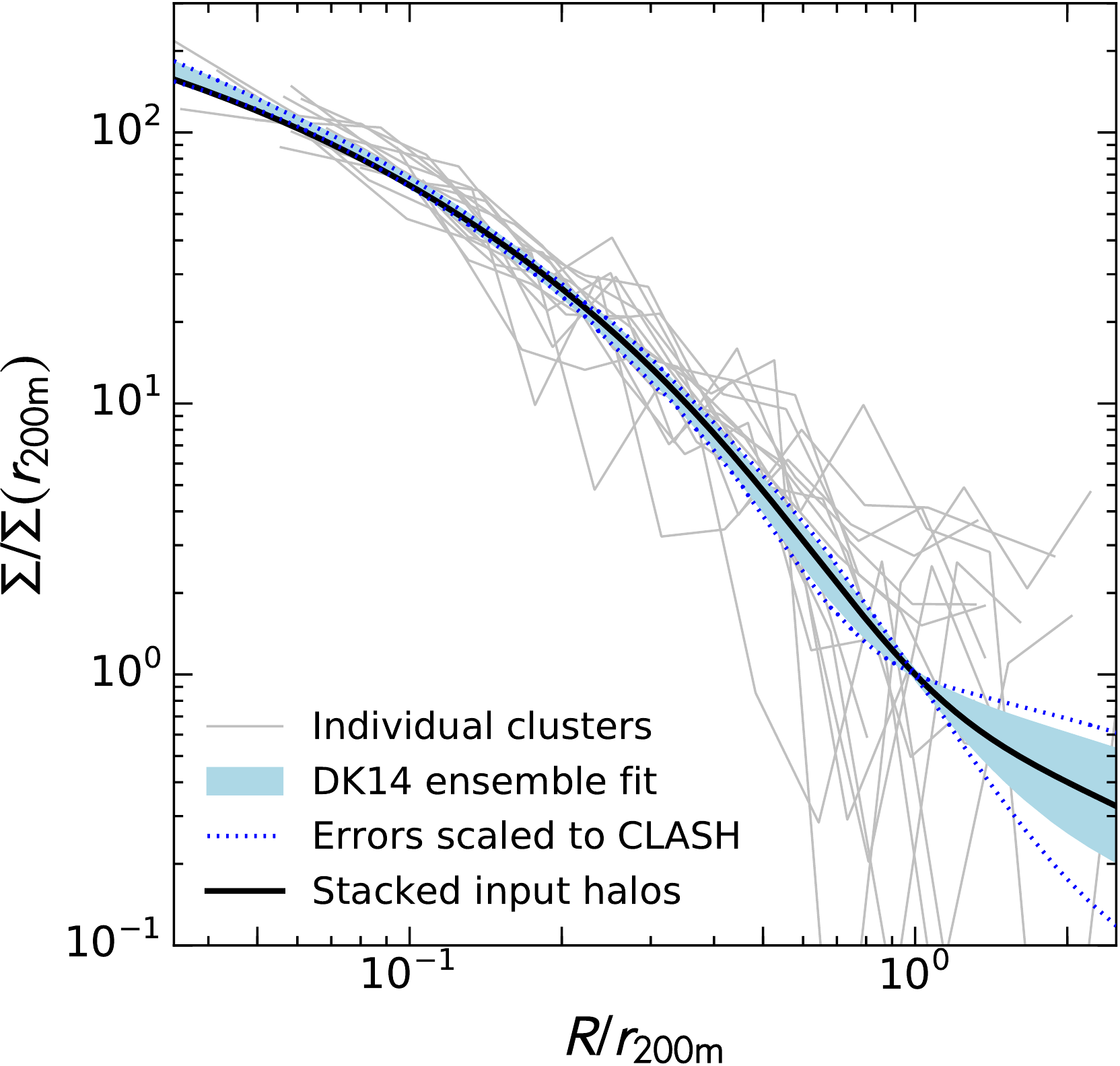} & 
    \includegraphics[scale=0.45, angle=0,clip]{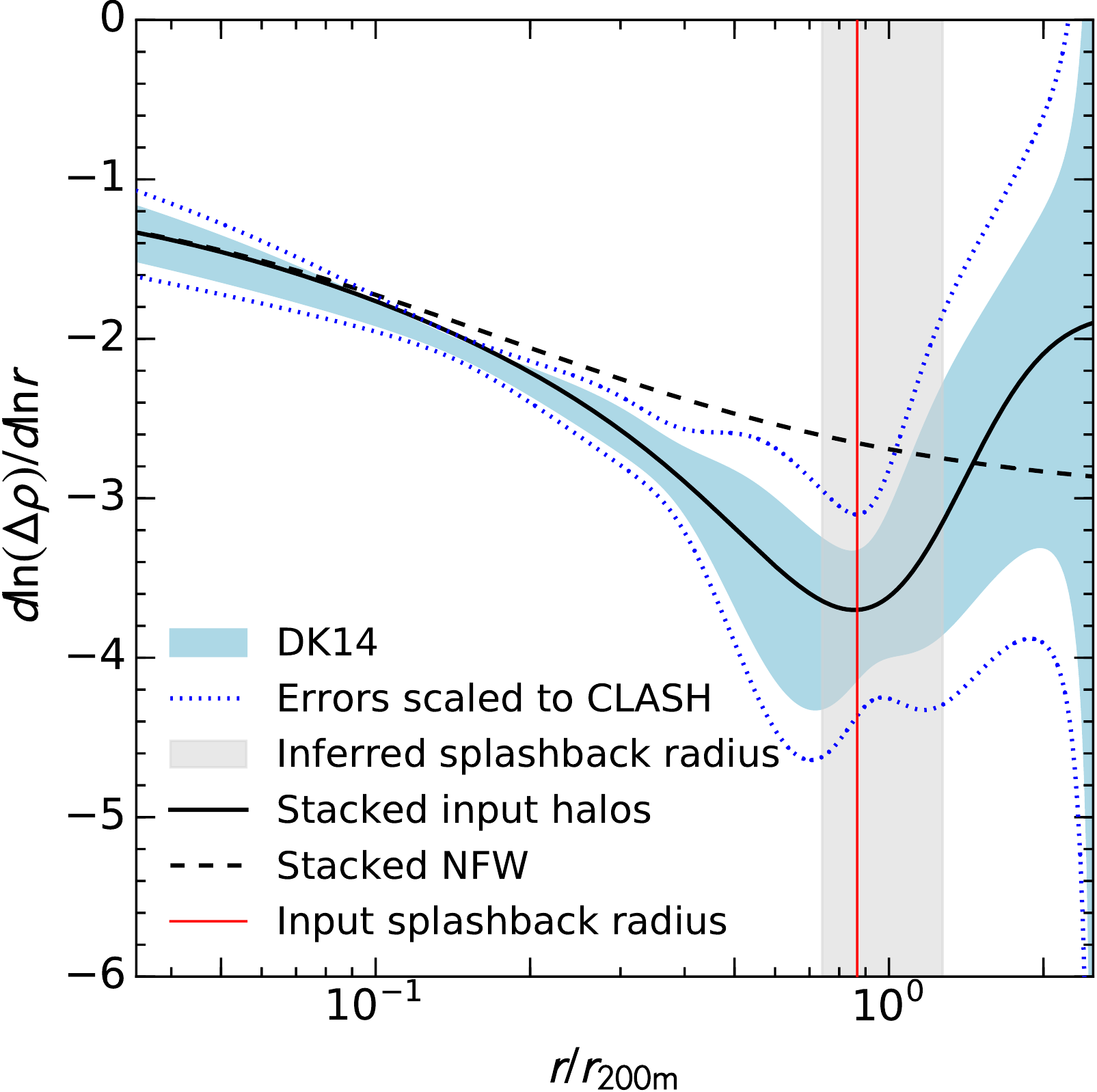}
  \end{array}
  $
 \end{center}
 \caption{\label{fig:mock}
 Test of our forward-modeling method. We create 50 realizations of 
 CLASH-like weak-lensing data by computing synthetic shear and
 magnification catalogs from analytically modeled cluster lenses.
 We simulate two configurations of the outer density profile,
 one without a splashback feature (top panels) and one with a
 splashback-like feature (bottom panels, see Appendix
 \ref{appendix:mock} for details on the lens models).
 The left panels show the scaled surface mass density 
 $\Sigma(R)/\Sigma(r_\mathrm{200m})$ of the synthetic observations, where
 the black solid line represents the noise-free, sensitivity-weighted
 profile of the 16 clusters in the synthetic sample.
 The blue shaded region indicates the $1\sigma$ bounds on the ensemble
 of DK14 fits to the synthetic data.
 The gray lines show scaled $\Sigma(R)$ profiles of individual
 clusters reconstructed from each particular source realization.
 The right panels show the logarithmic slope of the three-dimensional 
 DK14 profiles, $d\ln\Delta\rho(r)/d\ln{r}$, as a function of 
 $r/r_\mathrm{200m}$. For the lens model with a splashback-like feature, 
 the range from the 16th to the 84th percentile of $\Rsp$
 (gray vertical shaded area) 
 inferred from the synthetic data is consistent with the
 sensitivity-weighted expectation value of the sample
 (red vertical line), $\langle \Rsp/r_\mathrm{200m}\rangle\simeq 0.87$.
 In each panel, the blue dotted lines indicate the errors scaled to
 match the overall CLASH weak-lensing sensitivity.
 This test demonstrates the robustness of our fitting method, and in
 particular that fitting the DK14 profile does not introduce a spurious
 splashback feature. See Section \ref{subsec:mock} and Appendix
 \ref{appendix:mock} for details.
}
\end{figure*}

\begin{deluxetable*}{ccccccccccc}
\tablecolumns{11}
\tablecaption{
\label{tab:models}
Best-fit NFW, Einasto, and DK14 parameters for the CLASH Sample
}
\tablewidth{0pt}
\centering
\tablehead{
\multicolumn{1}{c}{$\Delta$} &
\multicolumn{2}{c}{NFW} &
\multicolumn{3}{c}{Einasto} &
\multicolumn{5}{c}{DK14}
\\
 \multicolumn{1}{c}{} &
 \multicolumn{1}{c}{$c_\Delta$} &
 \multicolumn{1}{c}{$\chi^2/\mathrm{dof}$} &
 \multicolumn{1}{c}{$c_\Delta$} &
 \multicolumn{1}{c}{$\alpha$} &
 \multicolumn{1}{c}{$\chi^2/\mathrm{dof}$} &
 \multicolumn{1}{c}{$c_\Delta$} &
 \multicolumn{1}{c}{$\alpha$} &
 \multicolumn{1}{c}{$\tau_\Delta$} &
 \multicolumn{1}{c}{$B_\Delta$} &
 \multicolumn{1}{c}{$\chi^2/\mathrm{dof}$} 
}
\startdata
200m   & $5.13 \pm 2.9\%$ & 181/235 & $4.57 \pm 4.9\%$ & $0.200 \pm 9.3\%$ & 181/234 & $4.96 \pm 25\%$ & $0.198 \pm 28\%$ & $>0.82$ ($1.13$) & $0.28 \pm 0.41$ & 180/232\\
virial & $4.58 \pm 2.9\%$ & 179/235 & $4.08 \pm 4.5\%$ & $0.205 \pm 9.6\%$ & 179/234 & $4.43 \pm 25\%$ & $0.208 \pm 26\%$ & $>0.93$ ($1.40$) & $0.20 \pm 0.37$ & 179/232\\
200c   & $3.66 \pm 3.1\%$ & 177/235 & $3.30 \pm 3.9\%$ & $0.210 \pm 11\%$  & 178/234 & $3.58 \pm 25\%$ & $0.214 \pm 26\%$ & $>1.17$ ($1.75$) & $0.15 \pm 0.31$ & 177/232\\
500c   & $2.38 \pm 3.5\%$ & 172/235 & $2.19 \pm 3.5\%$ & $0.208 \pm 13\%$  & 171/234 & $2.43 \pm 27\%$ & $0.218 \pm 25\%$ & $>1.85$ ($2.92$) & $0.10 \pm 0.20$ & 171/232\\
2500c  & $1.05 \pm 5.2\%$ & 153/235 & $0.85 \pm 14\%$  & $0.174 \pm 22\%$  & 151/234 & $1.06 \pm 33\%$ & $0.182 \pm 34\%$ & $>3.12$ ($4.68$) & $0.15 \pm 0.13$ & 151/232
\enddata
\tablecomments{
The global best-fit parameters and their $1\sigma$ fractional errors (in per cent) for each model, with five different values of the scaling overdensity $\Delta$. For the DK14 parameter $B_\Delta$, we give the best-fit parameter and its $1\sigma$ uncertainty. For the DK14 parameter $\tau_\Delta$, we provide lower limits (68\% CL) and best-fit parameter values in parentheses. 
}  
\end{deluxetable*}

Given the simulation results of DK14, we expect the signature of the
splashback radius in lensing data to be weak. Thus, one concern is that
fitting with the DK14 profile function might introduce a spurious
``detection'' of a splashback feature due to systematics or overfitting
in the presence of noise. In order to address this potential issue, we
have tested our procedure (including data analysis, mass reconstruction,
stacking, and the fitting process) on simulated lensing data. 
We focus on the recovery of
the lensing signal in the noisy outer regions around $r_\mathrm{200m}$
where the splashback feature is expected. Hence, we consider only the
wide-field weak-lensing observables, namely the shear and magnification
effects in the subcritical regime. To this end, we create 50 source 
realizations of synthetic shear and
magnification catalogs for our 16 CLASH clusters,
each modeled as an NFW  halo specified by its redshift $z_\mathrm{l}$ 
(Table \ref{tab:sample}) and ($M_\mathrm{200c},c_\mathrm{200c}$)
parameters fixed to the observed central values (Table 2
of U16).
For each NFW cluster, we consider two configurations of the outer density
profile, one with and one without a splashback-like feature. For technical 
reasons, we do not use a DK14 profile for the synthetic cluster lenses
and instead substitute a profile that introduces a similar density 
drop (see Appendix \ref{appendix:mock} for details).

The results of applying our methods to the synthetic weak-lensing data
are summarized in Figure \ref{fig:mock}. The lower and upper panels
correspond to the simulations with and without a splashback-like feature,
respectively. The blue shaded region in the left panels shows the mean
and standard deviation of the best-fit DK14 profiles inferred from 50
realizations of the synthetic data for each configuration.
On average, this fit is in excellent agreement with the noise-free,
sensitivity-weighted averaged input profile (black solid curve).
In the right panels, we show the corresponding logarithmic
density slope $d\ln\Delta\rho(r)/d\ln{r}$.
For the model with a splashback-like feature, the range from
the 16th to the 84th percentile of $\Rsp$
($0.74 \le \Rsp/r_\mathrm{200m} \le 1.27$; vertical shaded area) inferred
from the synthetic data is consistent with the sensitivity-weighted
expectation value of the sample,  
$\langle \Rsp/r_\mathrm{200m}\rangle\simeq 0.87$.

In Figure \ref{fig:mock}, we also indicate the errors scaled to the 
CLASH weak-lensing sensitivity (dotted lines). The errors are computed
by matching the synthetic (NFW) to the observed (CLASH) total
signal-to-noise ratio (S/N) of the $\bSigma$ profiles, where
$(\mathrm{S/N})^2=\sum_{n=1}^{N_\mathrm{halo}}\left(\bSigma^t C^{-1}\bSigma\right)_n$.
This comparison suggests that the overall uncertainty in our synthetic
observations is underestimated by $\simeq 36\%$.\footnote{The
underestimation is partly due to the assumed ellipticity dispersion of
source galaxies, $\sigma_g=0.3$ (Appendix \ref{appendix:mock}).
This is $\simeq 29\%$ lower than the
observed value, $\sigma_g\simeq 042$, which includes contributions fro
both intrinsic shape and measurement noise. The rest ($\sim 20\%$) can be
accounted for by the intrinsic clustering and other error contributions
in the weak-lensing magnification measurements
\citep{Umetsu2014clash,Umetsu2016clash} as well as by the cosmic noise
contribution, $C^\mathrm{lss}$.}
Nevertheless, the $1\sigma$ uncertainty after this correction is small
compared to the absolute value of the slope at $r \simlt 2r_\mathrm{200m}$.
Even though the determination becomes noisy at radii around and beyond
$r_\mathrm{200m}$,
the steepening relative to the NFW profile is marginally identified at
the $1.7\sigma$ level at the expected location, $r/r_\mathrm{200m}\sim 0.9$.


This test demonstrates that our analysis methods are able to
accurately reproduce the input sensitivity-weighted density profile and 
its logarithmic gradient even at realistic noise levels.
The results also show that the priors adopted (Section
\ref{subsec:bayesian}) are generic and flexible enough to reproduce the
NFW-like shape of the profile, as well as a splashback feature.
Importantly, we note that our analysis pipeline does not introduce spurious
gradients that mimic the characteristic splashback feature, 
namely a steepening followed by an upturn due to the contribution from
the 2-halo term.


\begin{figure*}[!htb] 
 \begin{center}
  $ 
  \begin{array}
   {c@{\hspace{.3in}}c}
    \includegraphics[scale=0.45, angle=0, clip]{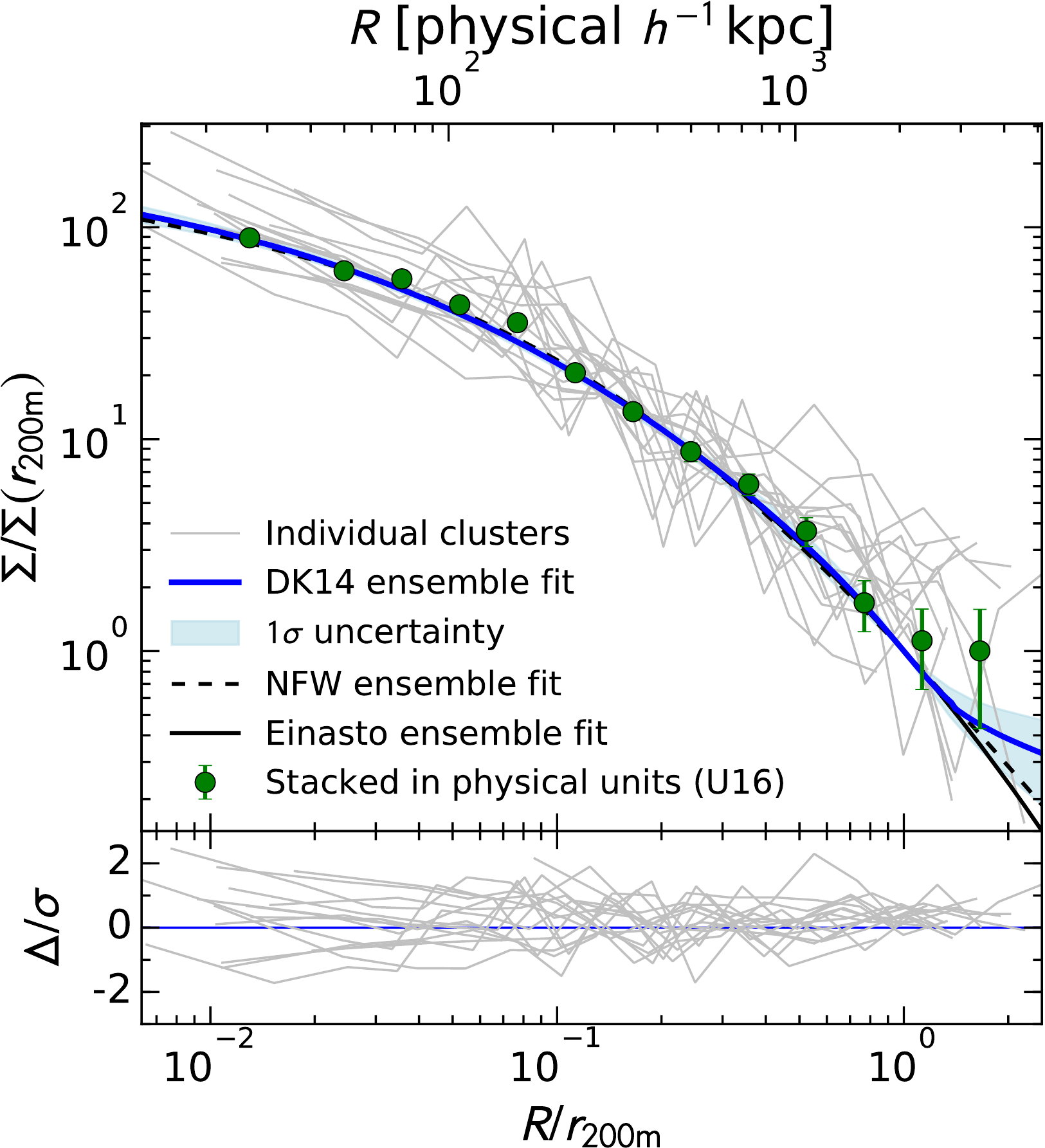} & 
    \includegraphics[scale=0.45, angle=0, clip]{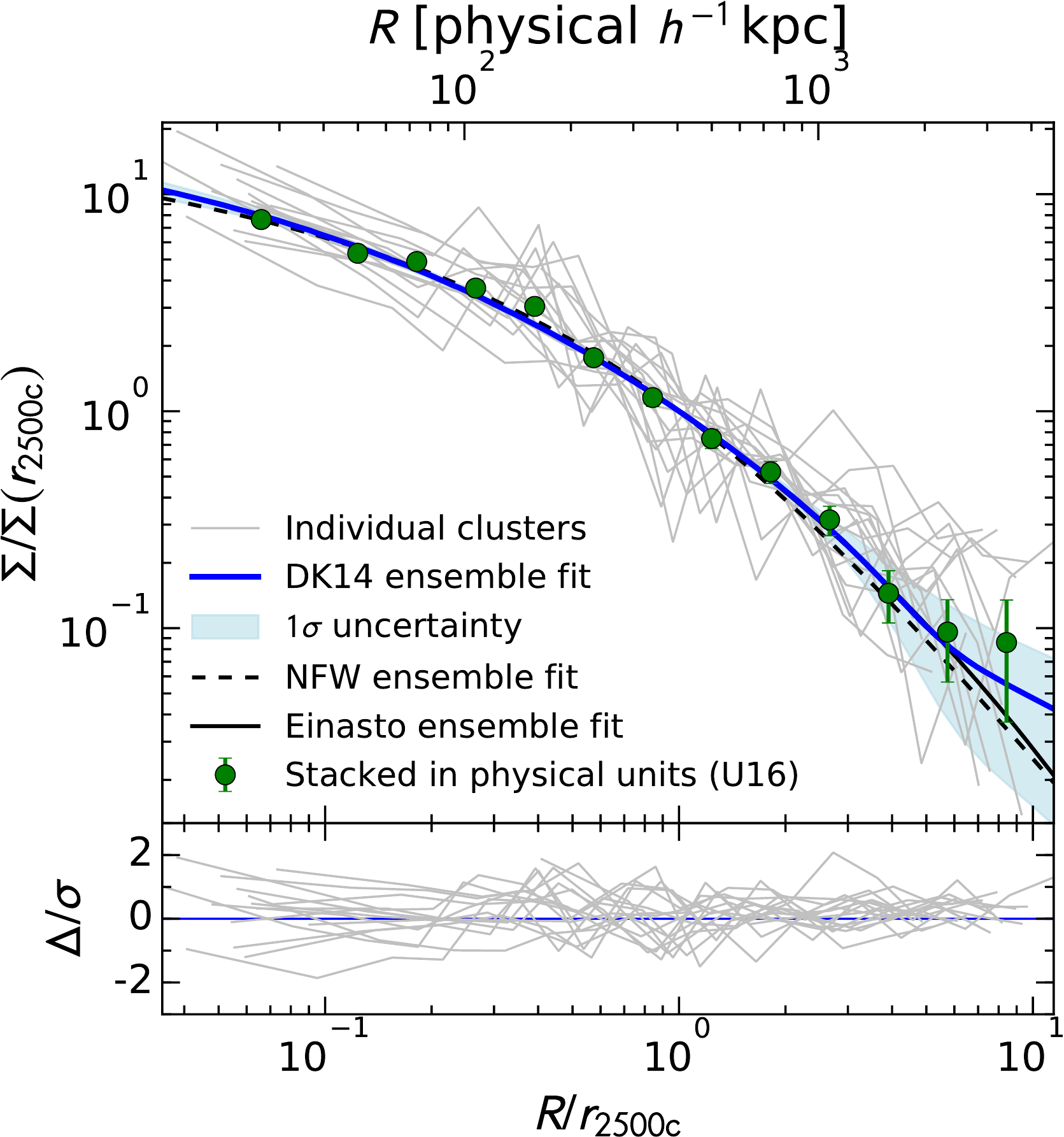}
  \end{array}
  $
 \end{center}
 \caption{
\label{fig:surfmass}
Upper panels: scaled surface mass density $\Sigma/\Sigma(r_\Delta)$ of
 the CLASH sample as a function of $R/r_\Delta$, where the projected
 clustercentric distance $R$ is expressed in units of two different 
 overdensity radii, $\Delta=\mathrm{200m}$ (left) and $\mathrm{2500c}$ (right).
 In each panel, the blue thick solid line and the blue shaded area show
 the best-fit DK14 profile and its $1\sigma$ uncertainty derived from a
 simultaneous ensemble fit to the scaled surface mass density 
 profiles of the 16 CLASH clusters (gray lines).
 The corresponding NFW (black dashed) and Einasto (black solid) fits
 are also shown.
 The scale on the top axis denotes $R$ in physical length units converted 
 with the effective overdensity radius $r_\Delta$ of the sample.
 The average $\Sigma$ profile of the CLASH sample stacked in physical
 units \citep[][U16]{Umetsu2016clash} is shown in rescaled units (green
 circles with error bars). 
 For each  $\Delta$, the lower panel shows deviations (in units of
 $\sigma$) of the observed cluster profiles from the best-fit DK14
 profile. 
}
\end{figure*}


\begin{figure*}[!htb] 
 \begin{center}
  $ 
  \begin{array}
   {c@{\hspace{.3in}}c}
    \includegraphics[scale=0.45, angle=0, clip]{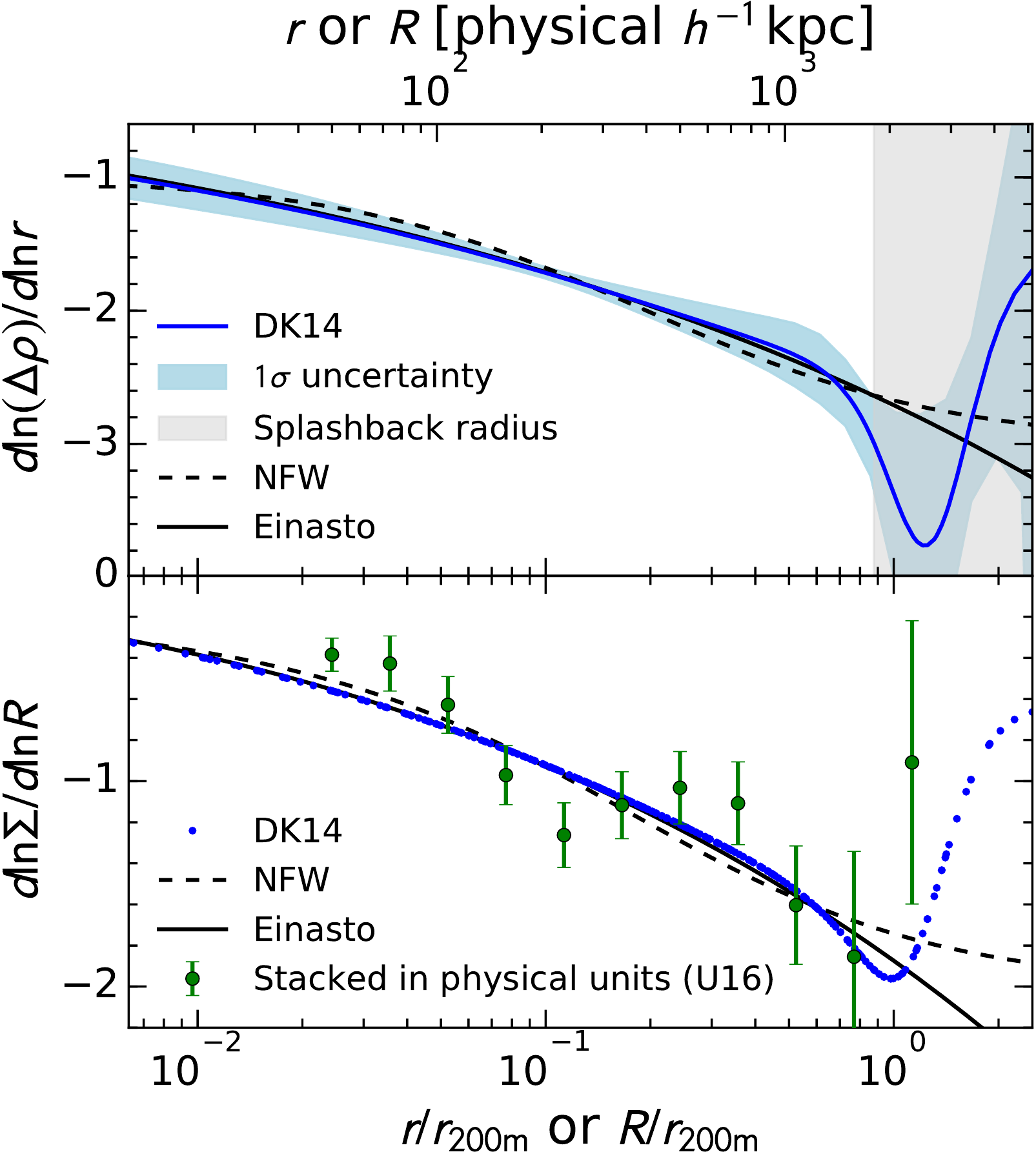} & 
    \includegraphics[scale=0.45, angle=0, clip]{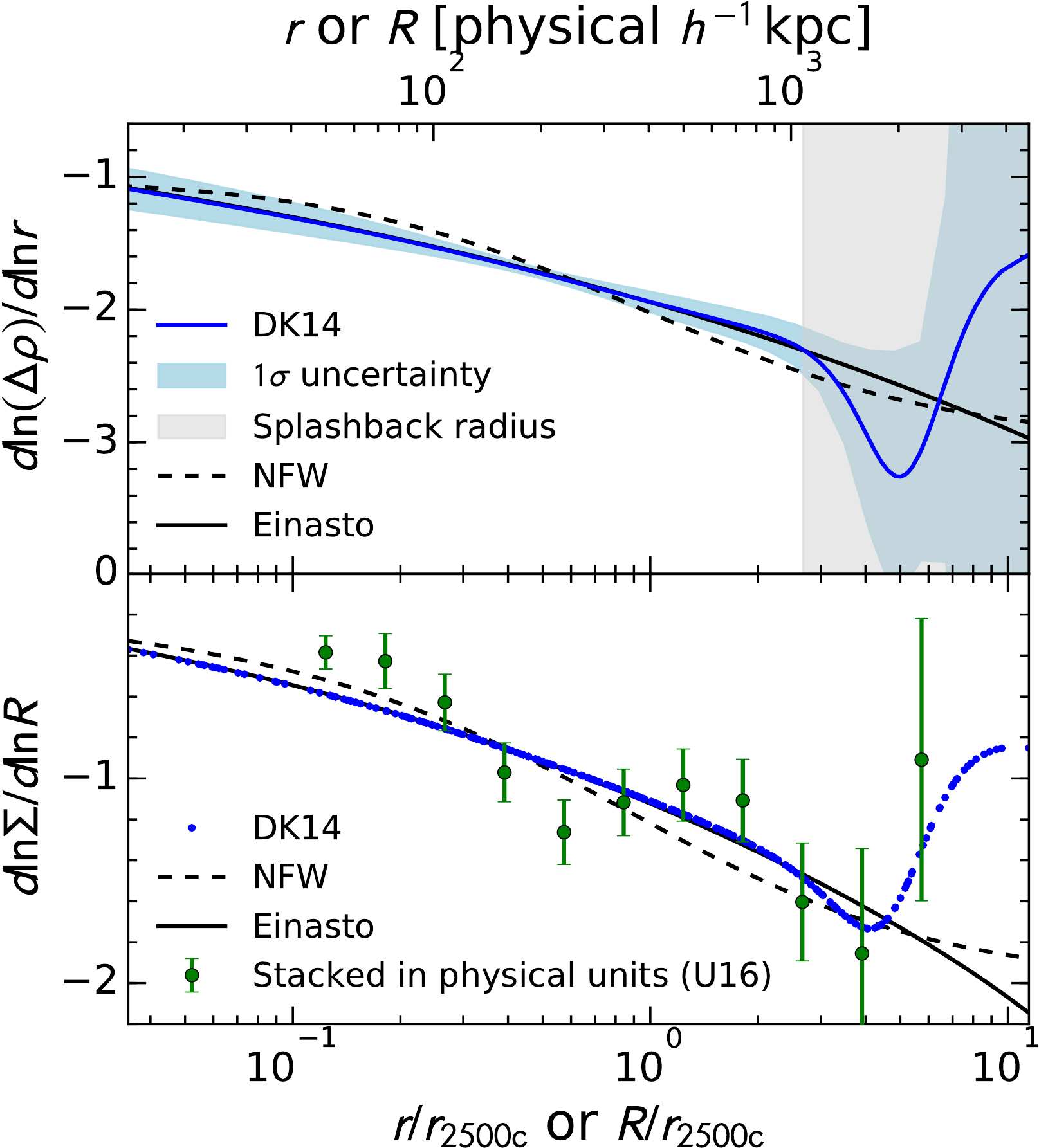}
  \end{array}
  $
 \end{center}
\caption{
\label{fig:Gamma3d}
 Same as Figure \ref{fig:surfmass}, but for the gradient of the profiles.
 Upper panels: logarithmic gradient of the inferred three-dimensional
 density profile as a function of the scaled cluster radius $r/r_\Delta$.
 As in Figure \ref{fig:surfmass}, the results are shown for
 $\Delta=\mathrm{200m}$ (left) and $\mathrm{2500c}$ (right). 
 The blue solid line and the blue shaded area represent the best-fit DK14
 model and its $1\sigma$ uncertainty, and are compared to the NFW (black
 dashed) and Einasto (black solid) fits.
 The gray vertical shaded area indicates the range from the
 16th to the 84th percentile
 of the marginalized posterior distribution of the splashback
 radius, $\Rsp/r_\Delta$ (see Figure \ref{fig:posterior}).
 Lower panels: same as the upper panels, but showing the logarithmic
 slope of the surface mass density profiles. The best-fit DK14 profile
 is shown as blue dots at the locations of the data points.
 For comparison, the slope of the conventionally stacked $\Sigma$
 profile (U16) is shown in rescaled units (green circles with error
 bars). 
}
\end{figure*}


\begin{figure*}[!htb] 
 \begin{center}
  \includegraphics[width=0.75\textwidth,angle=0,clip]{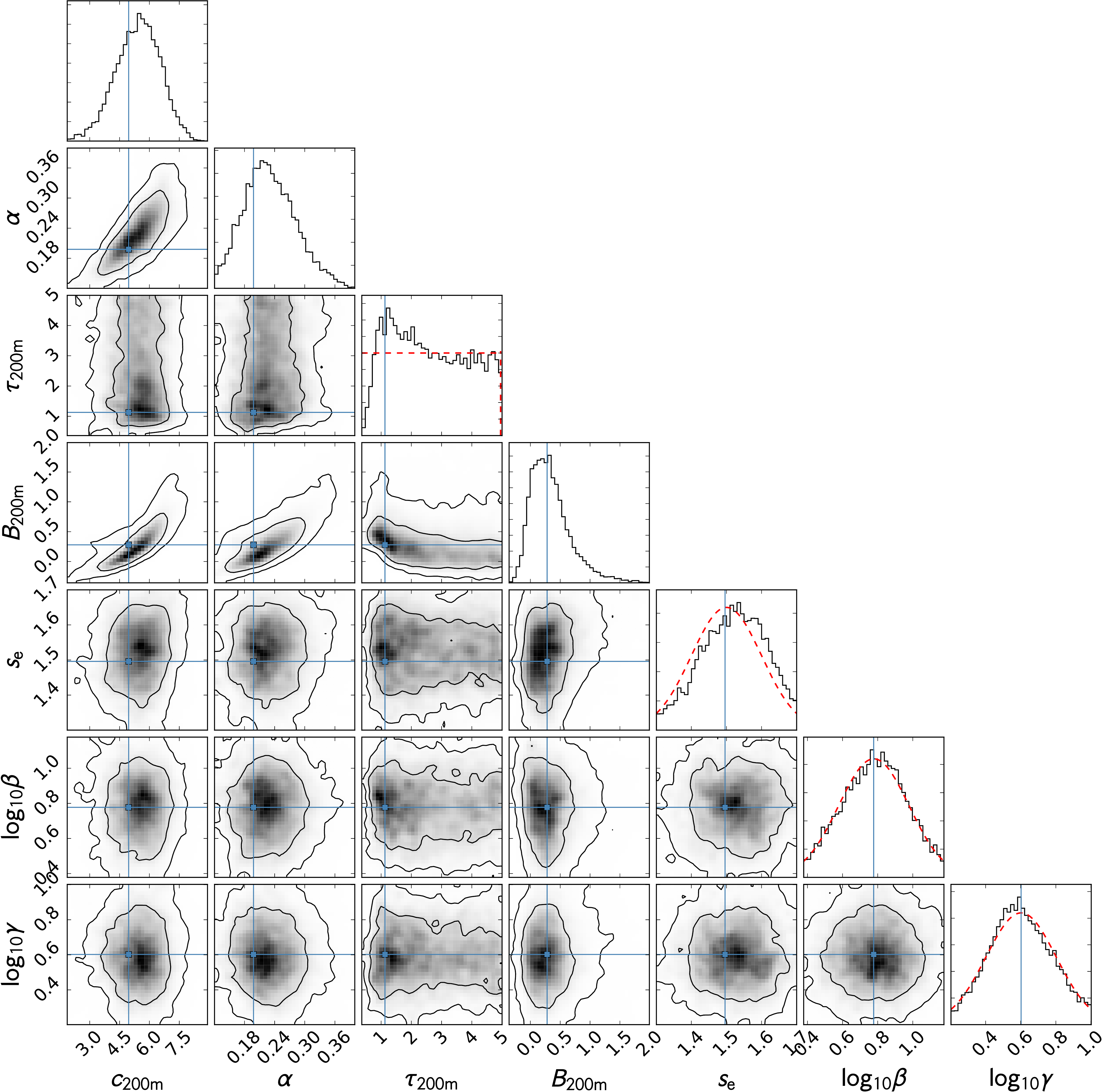} 
 \end{center}
 \caption{
\label{fig:DK14fit}
Constraints on the seven dimensionless parameters
$\bp=\{c_\mathrm{200m},\alpha,\tau_\mathrm{200m},B_\mathrm{200m},s_\mathrm{e},\beta,\gamma\}$
of the scaled DK14 model obtained from a simultaneous fit to
 the surface mass density profiles
 of 16 CLASH clusters (left panel of Figure \ref{fig:surfmass}), showing
 marginalized posterior one-dimensional distributions and
 two-dimensional 68\% and 95\% limits.   
Blue solid lines indicate the best-fit (global maximum of the posterior)
 values of the parameters.
For all parameters, the global best-fit values coincide well with their 
 respective peak values of the marginalized distributions. 
For ($\tau_\mathrm{200m},s_\mathrm{e},\log_{10}\beta,\log_{10}\gamma$), 
 the prior distributions are shown by red dashed lines. For the other 
 parameters, the priors are flat and nonrestrictive. 
}
\end{figure*}

\section{Results}
\label{sec:results}

Our main results are the best-fit NFW, Einasto, and DK14 profiles
resulting from a simultaneous fit to the scaled $\bSigma$ profiles of 16
CLASH clusters (Table \ref{tab:models} and Figures \ref{fig:surfmass}, 
\ref{fig:Gamma3d}, and \ref{fig:DK14fit}). Table \ref{tab:models} lists
the best-fit parameters, their uncertainties, and the $\chi^2$ values
for each of five pivot overdensities with which the analysis was
performed ($\Delta=\mathrm{2500c}, \mathrm{500c}, \mathrm{200c}$,
virial, and $\mathrm{200m}$).   

In Figure \ref{fig:surfmass}, we show the projected density profile
$y_\Delta(x)=\Sigma(r_\Delta x)/\Sigma(r_\Delta)$ of the individual
CLASH clusters (gray lines), as well as the best-fit DK14 (blue), NFW
(dashed black), and Einasto (solid black) fits. The results are shown
for two different choices of the pivot overdensity, namely 200m (left
panel) and 2500c (right panel). In the upper panels of Figure
\ref{fig:Gamma3d}, we show the corresponding three-dimensional logarithmic slope
$d\ln{\Delta\rho(r)/d\ln{r}}$ of the DK14 fit with $1\sigma$ errors
(shaded area), as well as the NFW and Einasto slopes.
Similarly, the lower panels show the logarithmic
slope $d\ln\Sigma(R)/d\ln{R}$ of the best-fit surface mass density
profiles. Finally, Figure \ref{fig:DK14fit} shows the one- and two-dimensional marginalized
posterior distributions for the complete set of scaled DK14
parameters with $\Delta=\mathrm{200m}$,
$\bp=\{c_\mathrm{200m},\alpha,\tau_\mathrm{200m},B_\mathrm{200m},s_\mathrm{e},\beta,\gamma\}$.
For all parameters, the global best-fit values coincide well with their respective peak values
of the one-dimensional marginalized posterior distributions. 

In the following sections, we discuss the fit quality as well as the
inferred parameters for the inner ($r \simlt \rvir$) and
outer regions of the profile.  

\subsection{Fit Quality}
\label{subsec:fit_quality}

The ensemble mass profile in projection is remarkably well described by 
an NFW or Einasto profile out to
$R\sim 1.2r_\mathrm{200m}$ 
or
$R\sim 4.5 r_\mathrm{2500c}$ (Figures \ref{fig:surfmass} and
\ref{fig:Gamma3d}),  
beyond which the data exhibit a flattening that is not modeled by those  
fitting functions. 
Note that to calculate $r_\Delta$ and $\Sigma(r_\Delta)$ for each 
individual cluster (Section \ref{subsec:data}), we employed the
spherical NFW fits of U16 obtained with a restricted fitting
range of $R\le 2\Mpch \sim r_\mathrm{200m}$. The results shown here thus
ensure the self-consistency of our analysis. 

As the outer profiles are expected to be most universal with respect to
$\Delta=\mathrm{200m}$ (DK14), that definition is of particular
relevance for the splashback radius.  We thus use the DK14 model with
$\Delta=\mathrm{200m}$ as a baseline model.  
This model has the best-fit $\chi^2$ of 180 for 232 degrees of freedom,
corresponding to a probability of 99.5\% to exceed the observed $\chi^2$
value, assuming the standard $\chi^2$ probability distribution
function. The model is therefore in good agreement with the data.
However, as we will discuss in Section
\ref{sec:discussion:profile_form}, the improvement in the fit is not
significant compared to the NFW or Einasto fit, implying that the parameters
that describe the transition region and outer terms are not well
constrained by the data.
This is not surprising, because Figure \ref{fig:surfmass} shows that the
CLASH lensing data do not resolve the profile curvature in the
transition region particularly well. Hence, the shape of the gradient
feature at $r\sim r_\mathrm{200m}$, which locally deviates from the
three-dimensional NFW and Einasto profiles (Figure \ref{fig:Gamma3d}),
is specific to the assumed DK14 profile form.

We note that the $\chi^2$ values in Table \ref{tab:models} decrease with   
increasing overdensity $\Delta$, independent of the fitting
function. The reason for this trend is that the inner $\bSigma$ profiles   
are more tightly constrained by the data, especially from the {\em HST}
lensing analysis, so that scaling the $\bSigma$ profiles to higher
overdensities reduces the overall scatter, which is dominated by the 
inner regions.
We also note that the reduced $\chi^2$ values in Table
\ref{tab:models} are systematically smaller than unity, which may
indicate that the number of degrees of freedom is overestimated owing to
the effects of nonlinear modeling \citep{Andrae2010} and/or that the
errors are conservatively overestimated. Since U16 found the reduced
$\chi^2$ for their fits to the same input data to be $\simgt 1$ (their Table 4),
it is unlikely that the errors are significantly overestimated.

\subsection{The Inner Mass Profile: Shape and Self-similarity}
\label{subsec:models}

The best-fit values of concentration shown in Table \ref{tab:models}
agree well between the different fitting models. This similarity is also
apparent in Figures \ref{fig:surfmass} and \ref{fig:Gamma3d}, as the
models have very similar shapes for both scaling overdensities
shown. Furthermore, the best-fit values for the shape parameter $\alpha$
agree between the Einasto and DK14 fits, and lie in the range  
$0.18\simlt \alpha\simlt 0.22$ with a typical $1\sigma$ uncertainty of
$0.06$ for the DK14 model, 
and $0.17\simlt\alpha\simlt 0.21$ for the Einasto model.
An Einasto density profile with $\alpha\sim 0.2$ closely resembles
an NFW profile \citep[e.g.,][]{Ludlow+2013}.

The uncertainties on $c$ and $\alpha$ allow us to assess the impact of
the scaling overdensity. If the profiles are most universal as a
function of a particular radius definition, we expect the fractional
uncertainty on the fit to be smallest in that definition. DK14
investigated the universality of halo density profiles, and found that
the inner profiles are most universal in units of
$r_\mathrm{200c}$. However, this statement refers primarily to the
redshift scaling of different definitions, which we cannot test here
owing to the limited redshift range of the CLASH sample. At fixed
redshift and fixed mass, we expect any overdensity within a range around
$r_\mathrm{200c}$ to lead to  reasonably universal inner profiles,
whereas for very extreme definitions the scatter in the profiles might
increase.   

These expectations are borne out in our results. The fractional
uncertainties on NFW-$c$ and Einasto-$\alpha$
(the primary parameter that determines the shape of the profile for each model) 
are smallest for the 200m and virial scalings, and increase toward the
highest overdensities (despite a lower $\chi^2$). 
However, the uncertainties on Einasto-$c$ are slightly lower at somewhat
higher overdensities such as $\Delta=500$c.
Overall, it appears that a rescaling with densities
around $\Delta = \mathrm{200c}$ leads to relatively low
uncertainties. Regardless of the profile model, scaling with
$\Delta=2500$c results in significantly higher uncertainties.
Since the determination of spherical overdensity radii (or masses) is
not less certain at high overdensities (U16, their Section 4.2), we
conclude that the increased uncertainty arises because the inner
profiles are less universal at very high overdensities.
Furthermore, we note that the relative insensitivity of the inner
profiles to $\Delta$ is likely in part due to the sample selection based
on X-ray regularity (Section \ref{subsec:data}), which is understood to
significantly reduce the scatter in concentration
\citep{Meneghetti2014clash}. The results have been confirmed by the
CLASH full lensing analysis of U16 (see their Section 6.2, as well as
\citealt{Merten2015clash}).  

Most importantly, we find that scaling the profiles by {\em any}  
overdensity radius (except for $\Delta = {\rm 2500c}$) improves the
constraints on the best-fit parameters. As a result, the limits on
concentration are tighter than those of U16 who scaled the profiles in
physical units and found uncertainties of 8\% on NFW-$c$, 11\% on
Einasto-$c$, and 18\% on Einasto-$\alpha$ (see their Table 4).


\begin{figure} 
 \begin{center}
 $ 
 \begin{array}
  {c@{\hspace{.1in}}c}
   \includegraphics[scale=0.25, angle=0, clip]{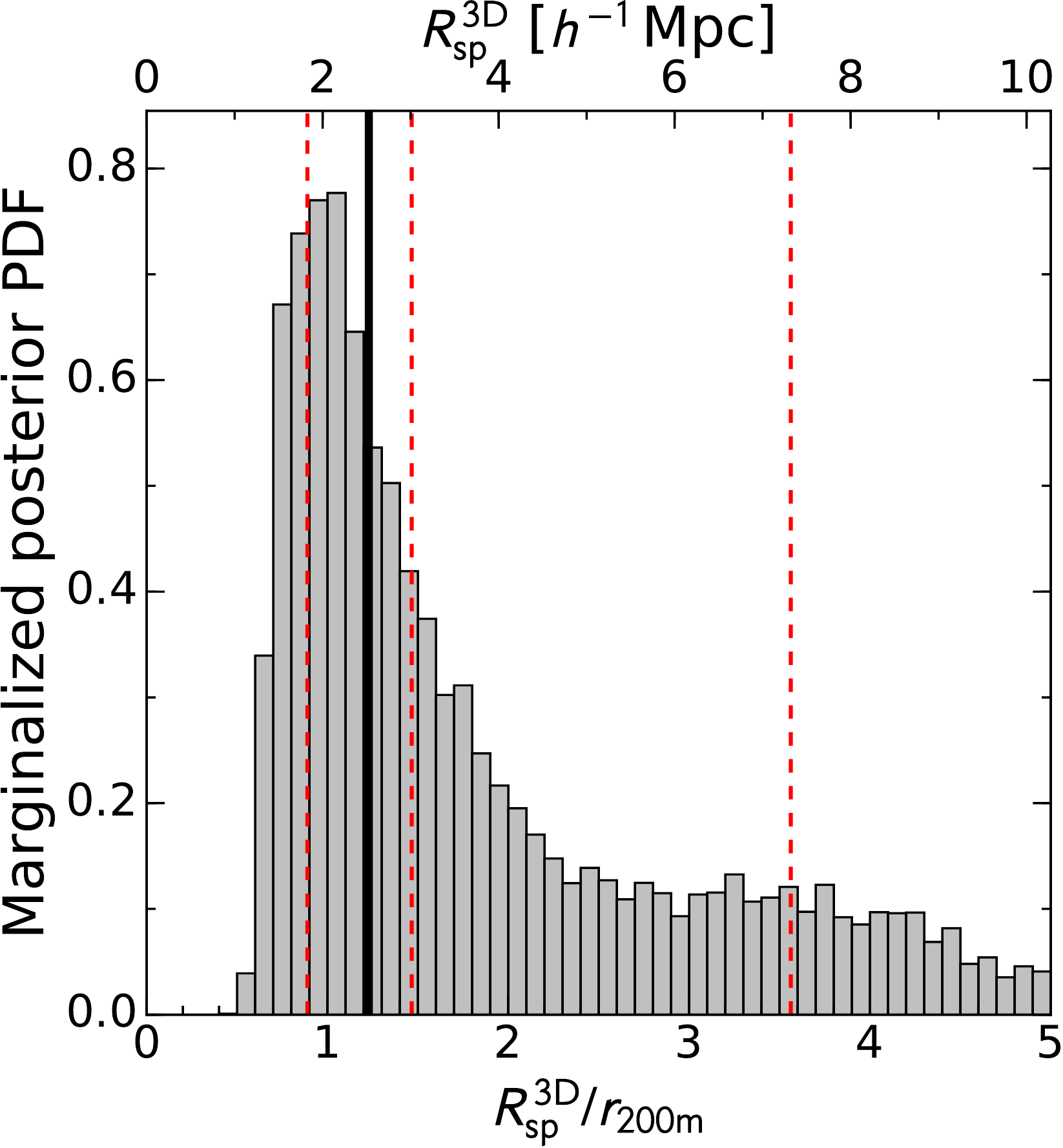} & 
   \includegraphics[scale=0.25, angle=0, clip]{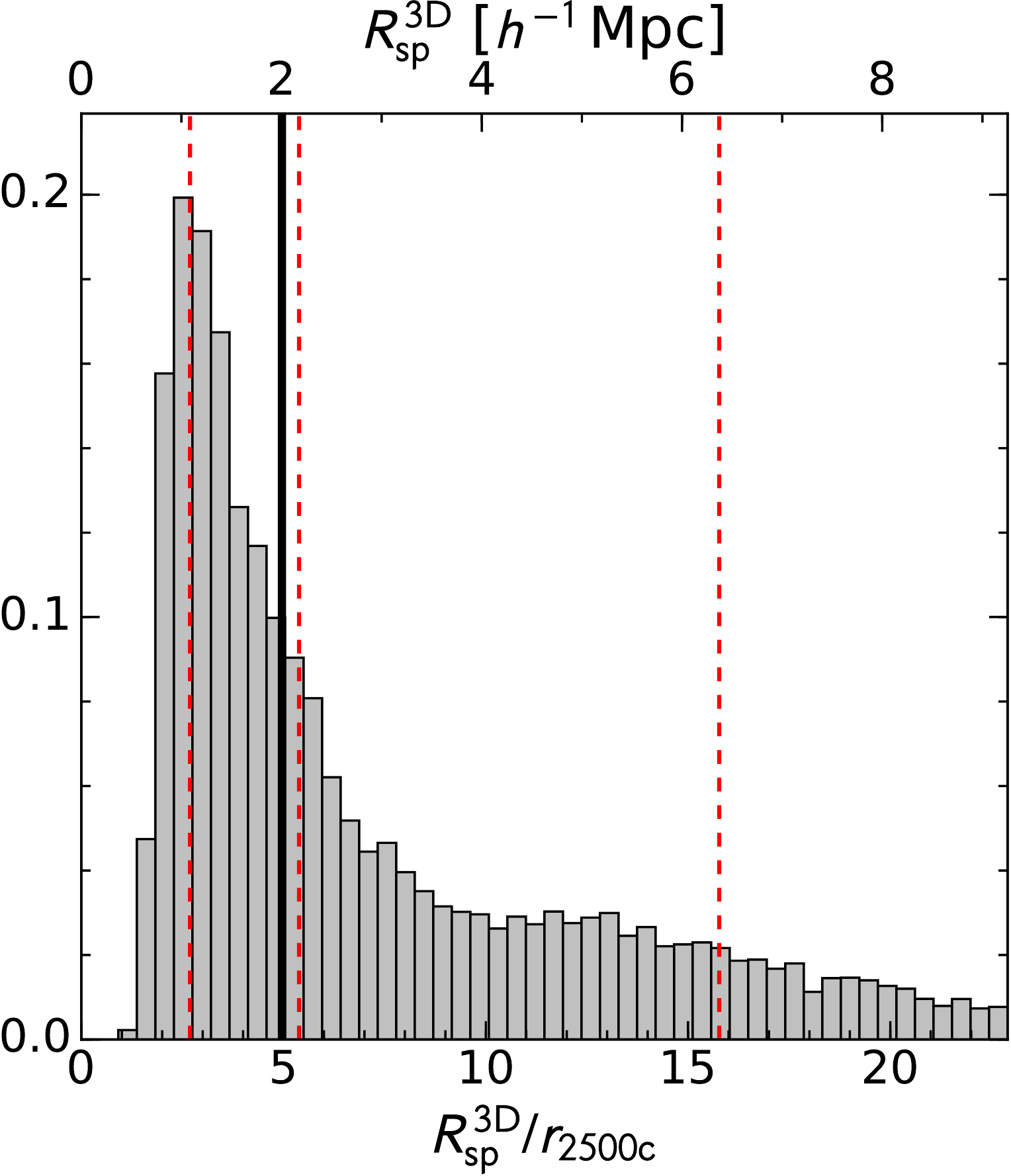}\\
   \includegraphics[scale=0.25, angle=0, clip]{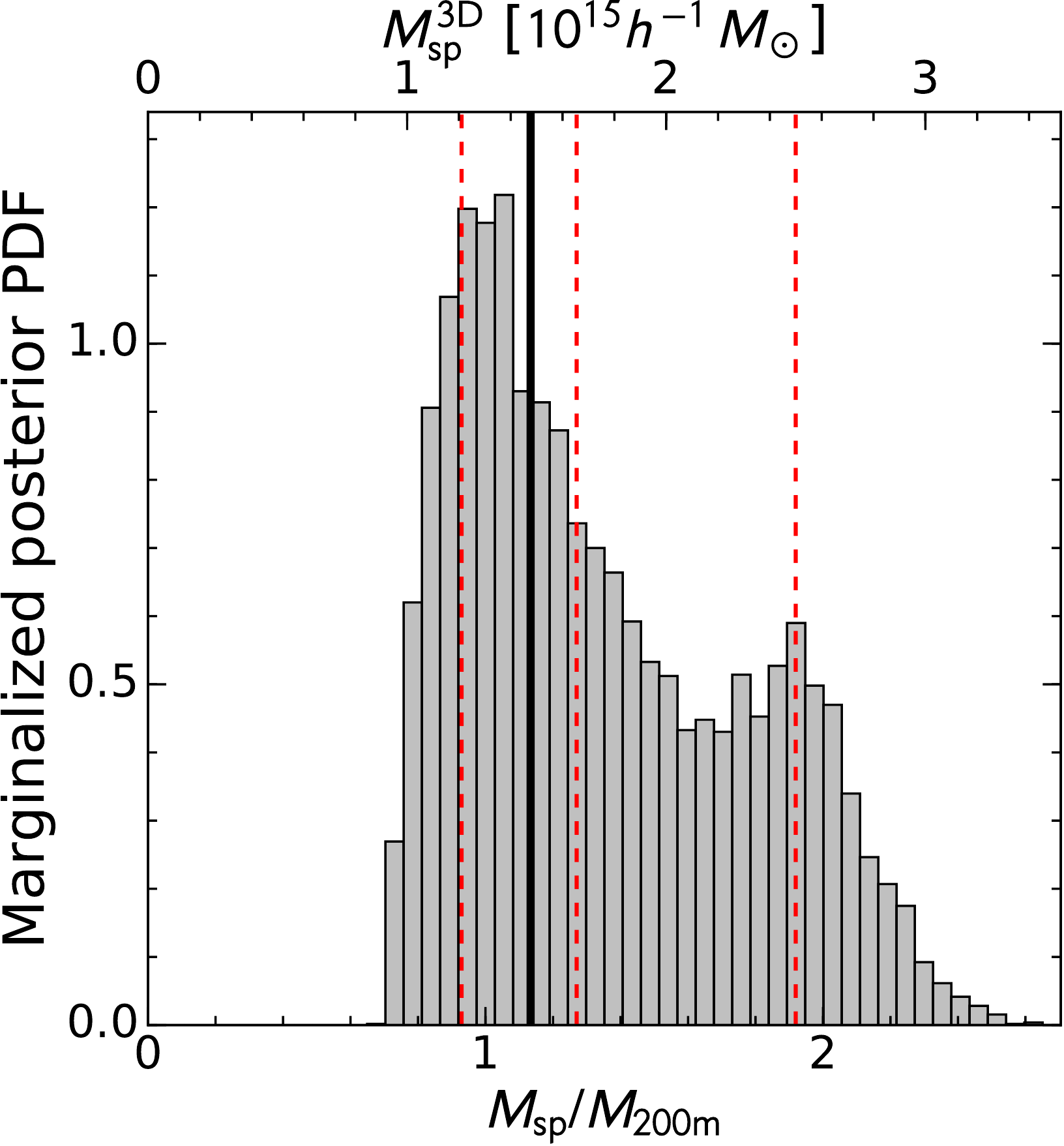} & 
   \includegraphics[scale=0.25, angle=0, clip]{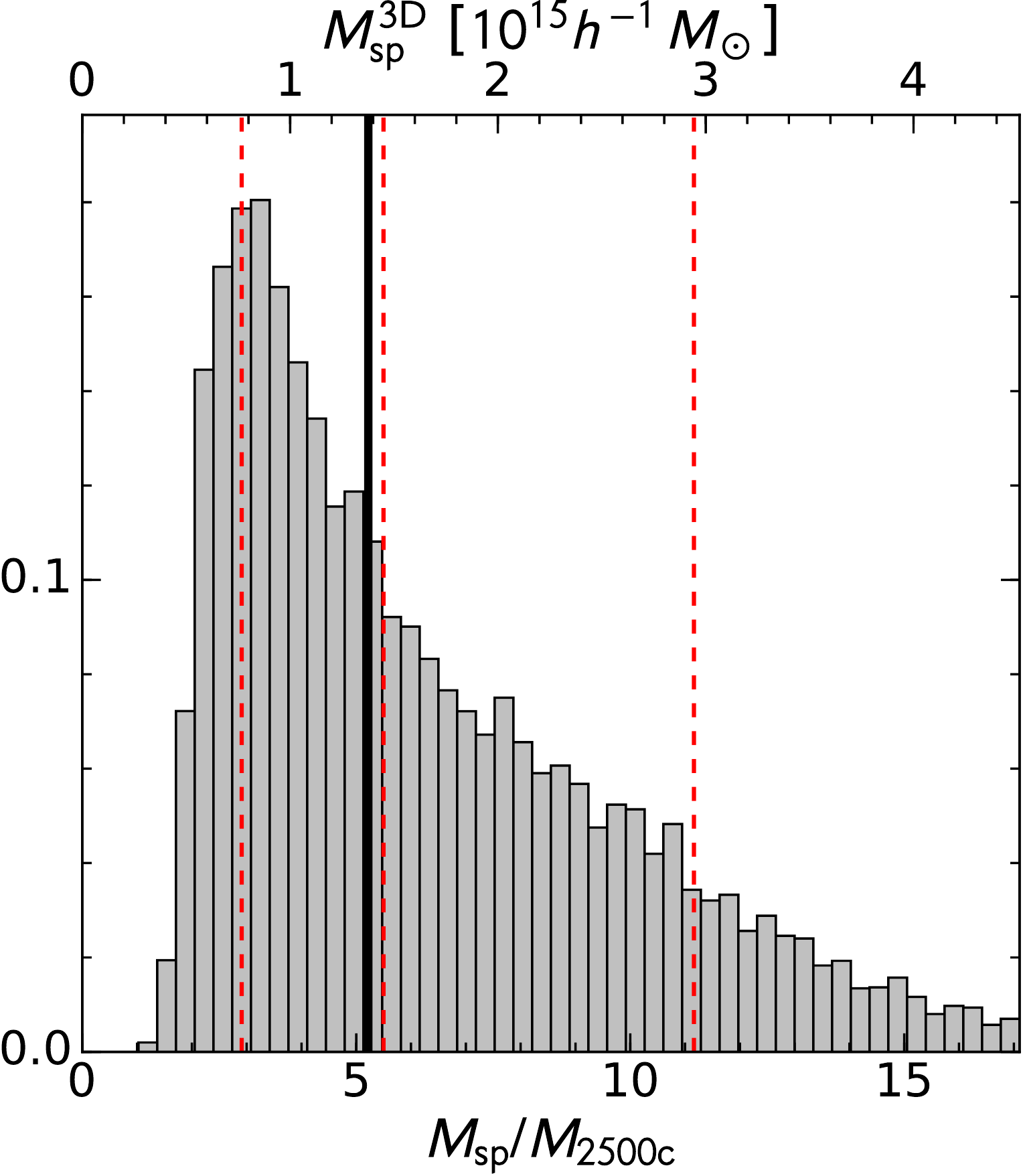}
  \end{array}
 $
 \end{center}
 \caption{\label{fig:posterior} 
Marginalized one-dimensional posterior probability distributions of the
 three-dimensional splashback radius (top panels) and mass (bottom
 panels) in scaled units, $\Rsp/r_\Delta$ and $M_\mathrm{sp}/M_\Delta$,
 for two different values of the chosen pivot overdensity,
 $\Delta=\mathrm{200m}$ (left panels) and $\mathrm{2500c}$ (right
 panels). The red vertical dashed lines indicate the
 16th, 50th, and 84th percentiles of the distributions. The global
 best-fit model values of $\Rsp/r_\Delta$ and $M_\mathrm{sp}/M_\Delta$
 are marked by a black vertical solid lines. The scales on the top axes
 denote $\Rsp$ and $M_\mathrm{sp}$ in physical units, converted using
 the effective overdensity radius $r_\Delta$ and mass $M_\Delta$ of the
 sample.   
}
\end{figure}

\subsection{The Outer Profile and Splashback Radius}
\label{subsec:Rsp}

We now turn toward the outer profiles and particularly the inferred
gradient profiles shown in Figure \ref{fig:Gamma3d}. A comparison of the
three- and two-dimensional slopes highlights why detecting the
splashback radius in surface density profiles is challenging in
practice: even though there is a noticeable steepening in the
three-dimensional slope, the two-dimensional slope drops very little,
owing to projection effects (see Section
\ref{sec:discussion:projection}).  

Nevertheless, we can derive a splashback radius and mass from each of
the MCMC samples of the DK14 parameters shown in Figure 
\ref{fig:DK14fit} (Equations \ref{eq:Rsp} and
\ref{eq:Msp}). In Figure 
\ref{fig:posterior}, we show the corresponding one-dimensional
marginalized posterior distributions for $\Rsp$ and $M_\mathrm{sp}\equiv
M(<\Rsp)$, both in scaled units.
The results are shown for two different values of the chosen pivot
overdensity, $\Delta=\mathrm{200m}$ (left panels) and 2500c (right panels).
In each panel, the vertical dashed lines indicate the 16th, 50th, and 84th
percentiles of the marginalized posterior distribution.
The locations of
$\Rsp/r_\Delta$ and
$M_\mathrm{sp}/M_\Delta$
derived from the best-fit model (Table \ref{tab:models})
are indicated by vertical solid lines in Figure \ref{fig:posterior}, and
are in close agreement with the respective median values of the
distributions.
The posterior distributions show a tail extending toward large positive 
values of $\Rsp/r_\Delta$ and $M_\mathrm{sp}/M_\Delta$,
associated with large values of the truncation parameter,
$\tau_\Delta=r_\mathrm{t}/r_\Delta$ (Figure \ref{fig:DK14fit}).
A large $\tau_\Delta$ indicates a profile without a well-defined
steepening feature. 
We thus place uninformative upper bounds on $\Rsp/r_\Delta$ and
$M_\mathrm{sp}/M_\Delta$.  
On the other hand, we obtain tighter lower bounds on these parameters 
because the inner $\bSigma$ profiles of the clusters are well
constrained by the combination of strong-lensing, weak-lensing shear and
magnification data (U16).

\begin{deluxetable}{ccccc}
\tablecolumns{5}
\tablecaption{
\label{tab:Rsp}
Constraints on the Splashback Radius and Mass
}
\tablewidth{0pt}
\centering
\tablehead{
 \multicolumn{1}{c}{$\Delta$} &
 \multicolumn{1}{c}{$R_\mathrm{sp}^\mathrm{3D}/r_\Delta$} &
 \multicolumn{1}{c}{$R_\mathrm{sp}^\mathrm{3D}$} &
 \multicolumn{1}{c}{$M_\mathrm{sp}/M_\Delta$} &
 \multicolumn{1}{c}{$M_\mathrm{sp}$}
\\ 
\multicolumn{1}{c}{} &
\multicolumn{1}{c}{} &
\multicolumn{1}{c}{$(\mathrm{Mpc}/h)$} &
\multicolumn{1}{c}{} &
\multicolumn{1}{c}{$(10^{15}M_\odot/h)$} 
}
\startdata
200m    & $>0.89$ ($1.23$) & $>1.83$ ($2.52$) & $>0.93$ ($1.13$) & $>1.21$ ($1.48$)\\
virial  & $>0.91$ ($1.52$) & $>1.67$ ($2.78$) & $>0.94$ ($1.30$) & $>1.13$ ($1.56$)\\
200c    & $>1.04$ ($1.90$) & $>1.52$ ($2.79$) & $>1.03$ ($1.55$) & $>1.04$ ($1.57$)\\ 
500c    & $>1.37$ ($3.09$) & $>1.30$ ($2.94$) & $>1.31$ ($2.39$) & $>0.90$ ($1.64$)\\
2500c   & $>2.69$ ($4.96$) & $>1.09$ ($2.00$) & $>2.91$ ($5.22$) & $>0.77$ ($1.38$)
\enddata
 \tablecomments{
Lower limits (68\% CL) and best-fit model values (in parentheses) for
 the three-dimensional splashback radius and mass.
 The splashback radius and mass in physical length units were
 converted using the effective overdensity radius
 $r_\Delta^\mathrm{eff}$ and mass $M_\Delta^\mathrm{eff}$ of the
 sample, respectively.
}
\end{deluxetable}

Table \ref{tab:Rsp} summarizes the 68\% confidence lower limits and
best-fit model values (in parentheses) on the splashback radius and
mass. We also list $\Rsp$ and $M_\mathrm{sp}$
in physical length units converted with the
effective overdensity radius ($r_\Delta^\mathrm{eff}$) and mass
($M_\Delta^\mathrm{eff}$) of the sample 
(Section \ref{subsec:data}).
Using the fiducial scaling overdensity of $\Delta=\mathrm{200m}$, these lower limits are
$\Rsp/r_\mathrm{200m}>0.89$
and
$M_\mathrm{sp}/M_\mathrm{200m}>0.93$,
corresponding to
$\Rsp > 1.83\Mpch$
and
$M_\mathrm{sp} > 1.21\times 10^{15}\Msunh$.
We note that the location of the steepest slope in projection,
$R_\mathrm{sp}^\mathrm{2D}/r_\Delta$, 
is smaller than $R_\mathrm{sp}^\mathrm{3D}/r_\Delta$ (Figure
\ref{fig:Gamma3d}) because of projection effects
\citep{More2016splash}. Using our best-fit base DK14 model
($\Delta=\mathrm{200m}$ in Table \ref{tab:models}), we find
$R_\mathrm{sp}^\mathrm{2D}/\Rsp\simeq0.8$. We
discuss the difference between the two- and three-dimensional splashback
radii further in Section \ref{sec:discussion:projection}. 
  
From a comparison of the lower bounds on the splashback radius and mass in
Table \ref{tab:Rsp}, we see that the steepening feature is most pronounced
when the cluster profiles are scaled by $r_\mathrm{200m}$, and is smeared
out when scaled to higher overdensities, resulting in less strict lower limits 
(see also the right panels of Figure \ref{fig:Gamma3d}).
This trend is consistent with the prediction of DK14 that
halos reveal self-similar behavior in their outskirts when their
profiles are expressed in units of spherical overdensity radii defined
with respect to the mean density of the universe, especially
$r_\mathrm{200m}$.


\section{Discussion}
\label{sec:discussion}

In this section, we compare our results for the shape of the profile, 
concentration, and splashback radius with predictions from $N$-body
simulations. We discuss observational and simulation
effects that could potentially complicate our analysis.

\subsection{The Impact of Priors}
\label{sec:discussion:tau}

We imposed a number of priors on the
parameters of the DK14 profile, namely
on the slopes $\beta$ and $\gamma$, on the steepening radius
$\tau_\Delta$, and on the outer profile slope $s_\mathrm{e}$ (see Figure
\ref{fig:DK14fit}).
These priors were based on the results of DK14 for the median profiles
of halo samples spanning a wide range of masses and mass accretion
rates, and chosen conservatively, i.e. allowing a much larger range of 
parameter values than found for high-mass cluster halos in DK14.
We find that imposing constraining theoretical priors leads 
to an inflated sensitivity to the splashback feature.
As stated in Section \ref{subsec:DK14}, we use the DK14 profile as a 
flexible fitting function to determine the location of the steepest
slope, $\Rsp$, which is an observable quantity.
In this context,
$\bp=\{c_\Delta,\alpha,\tau_\Delta,B_\Delta,s_\mathrm{e},\beta,\gamma\}$
are considered to be merely fitting parameters, and we allow them to take
on values not expected from simulations, such as very large $\alpha$.

Nevertheless, one might worry that our inferences regarding $\Rsp$ 
are informed by the priors because they clearly inform the posterior of
some parameters (Figure \ref{fig:DK14fit}).
In particular, the asymptotic slope of the 1-halo term, $\gamma$,
is essentially unconstrained by the fit and relatively steep due to the
prior (for example, $\gamma = 0$ is effectively excluded). However, it
is important to note that $\gamma$ (as well as $\beta$) has no impact on
the DK14 profile if $\tau_\Delta$ is large, because the steepening then
moves out of the observed region of the profile. Thus, the most critical
prior is that on $\tau_\Delta$. 

Our prior allows values of $\tau_\Delta$ up to 5, placing the steepening
at $r_\mathrm{t}\simlt 10\Mpch$ which lies far outside the maximum
radius of our data ($R\simeq 5\Mpch$).
Thus, profiles with $\tau_\mathrm{200m} \simgt 2.5$ effectively reduce
to an Einasto profile with a 2-halo term ($f_\mathrm{trans}\approx 1$) 
in the observed radial range.
We have already demonstrated that a fit with these priors can reproduce
a profile without steepening in the analysis of synthetic weak-lensing
data (Figure \ref{fig:mock}). 
However, we note that, in combination with a higher value of $\alpha$
and a positive value of $B_\Delta$, even profiles with
$\tau_\mathrm{200m}>2.5$ can reproduce a splashback feature in 
the observed radial range (with an inner profile steeper than the best-fit 
NFW/Einasto profile). We have confirmed that the resulting profiles 
are, in fact, a reasonable description of the data. Our priors also allow 
a profile with a negative 2-halo normalization,
$B_\Delta\le 0$ (i.e., underdense regions), which can produce a
steepening gradient without an upturn feature. Here, the ensemble fits
yield positive $\Delta\rho$ (i.e., $\rho>\rho_\mathrm{m}$) in the
observed range because $\bSigma$ is constrained to be positive (U16). 

In order to confirm that the $\tau_\Delta$ prior does not significantly
affect our results, we have performed a fit with a flat prior of $\tau_\mathrm{200m}<20$,
corresponding to $r_\mathrm{t}\simlt 40\Mpch$ for the CLASH sample. With
this relaxed $\tau_\Delta$ prior, we find the same best-fit solution and
a 68\% CL lower bound of $\Rsp/r_\mathrm{200m}>0.90$, compared to
$>0.89$ obtained with $\tau_\mathrm{200m}<5$.
This difference represents a $1\%$ change, which is not significant
given the current sensitivity. Moreover, we find no noticeable changes
in the constraints on the density and gradient profiles shown in Figures
\ref{fig:surfmass} and \ref{fig:Gamma3d}. Therefore, we conclude that
our results are sufficiently robust against the choice of the
$\tau_\Delta$ prior.

\subsection{Which Density Profile Does the CLASH Sample Prefer?}
\label{sec:discussion:profile_form}

A robust outcome of our analysis is that, regardless of the
pivot overdensity $\Delta$ chosen, the ensemble CLASH mass profile in
projection is in full agreement with the NFW or Einasto profile out to
$\sim r_\mathrm{200m}$,
consistent with previous lensing results \citep[e.g.,][]{Umetsu+2011stack,Umetsu2014clash,Umetsu2016clash,Newman+2013a,Okabe+2013,Niikura2015,Okabe+Smith2016}.
This also ensures the self-consistency of our analysis, in which we used 
the NFW fits of U16 to calculate $r_\Delta$ and $\Sigma(r_\Delta)$ for
individual clusters, where the fitting range was restricted to
$R\le 2\Mpch \sim r_\mathrm{200m}$.

Our base DK14 model with $\Delta=\mathrm{200m}$ gives a slightly better
fit than the corresponding NFW or Einasto model in terms of the best-fit
$\chi^2$ values (Table \ref{tab:models}).
The relative improvements are
$\Delta\chi^2=\chi^2_\mathrm{NFW}-\chi^2_\mathrm{DK14}\simeq 1$
and
$\Delta\chi^2=\chi^2_\mathrm{Einasto}-\chi^2_\mathrm{DK14}\simeq 1$
for three and two additional free parameters, respectively.
Hence, the improvement is not statistically significant, implying that
our inference of the outskirt feature depends on the choice of the
fitting function and priors.

As discussed in Section \ref{subsec:DK14} and by \citet{More2016splash},
we would apply two requirements to claim a detection of the splashback radius
with a DK14 model, namely (1) that the location of the steepest slope in three
dimensions with respect to $r_\Delta$ can be identified at high
significance and
(2) that this steepening is greater than that expected from a DK14 model with
$f_\mathrm{trans} = 1$ (i.e., $\tau_\Delta\gg 1$, reducing to an Einasto profile). 
The second criterion is important to ensure that the steepening is
actually associated with a density caustic rather than the transition to
the 2-halo term. 

Given these criteria, we do not find sufficient evidence for the
existence of a splashback feature in the CLASH lensing data
because the data do not have the 
sensitivity necessary to resolve the profile curvature in the transition region.
This result is not surprising, as demonstrated in our simulated
experiment (Section \ref{subsec:mock}).
On the other hand,
assuming the DK14 profile form and generic priors calibrated
with numerical simulations, we have placed lower 
limits on the splashback radius of the CLASH clusters (Table
\ref{tab:Rsp}), if it exists. Since we cannot rule out models with $f_\mathrm{trans}=1$, it is
possible that the observed gradient feature in the outskirts is a
statistical fluctuation. In Section \ref{subsec:mock}, we showed that
our analysis pipeline produces unbiased results in terms of both
$\bSigma$ profiles and ensemble DK14 fits, and does not create spurious
gradient features. Hence, the inferred outskirt feature is unlikely to
arise from systematic errors. 

An additional source of uncertainties in our scaling analysis is
the statistical errors on the NFW parameters of individual clusters,
which propagate into uncertainties in their scaling radius $R_\Delta$
and scaling density $\Sigma(R_\Delta)$. In this work, these scaling
parameters of individual clusters are fixed to their best-fit values
from the NFW fits of U16. Hence, although our scaling analysis has led
to significant improvements in the constraints on the inner profiles
relative to the conventional stacking (see Section
\ref{sec:discussion:sims}), these errors are likely to have smeared out 
local features to some level \citep{Niikura2015}. The degree of smearing
can be assessed by marginalizing over uncertainties in the NFW
parameters for all clusters, and such effects need to be accounted for in
future studies with a larger statistical sample of clusters and with a
higher statistical sensitivity.


\begin{figure*}[!htb] 
 \begin{center}
  $ 
  \begin{array}
   {c@{\hspace{.3in}}c}
    \includegraphics[scale=0.45, angle=0, clip]{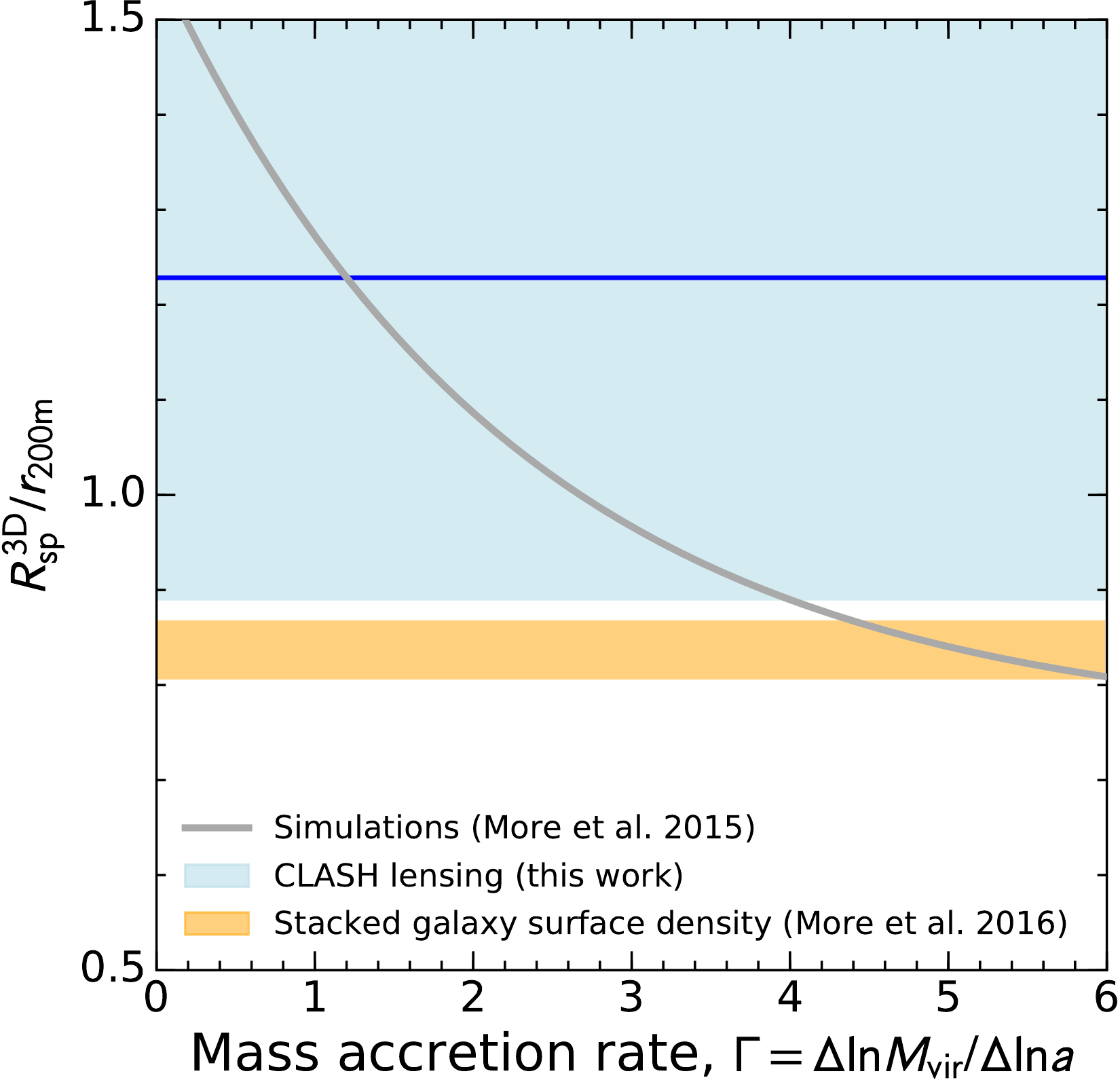} &
    \includegraphics[scale=0.45, angle=0, clip]{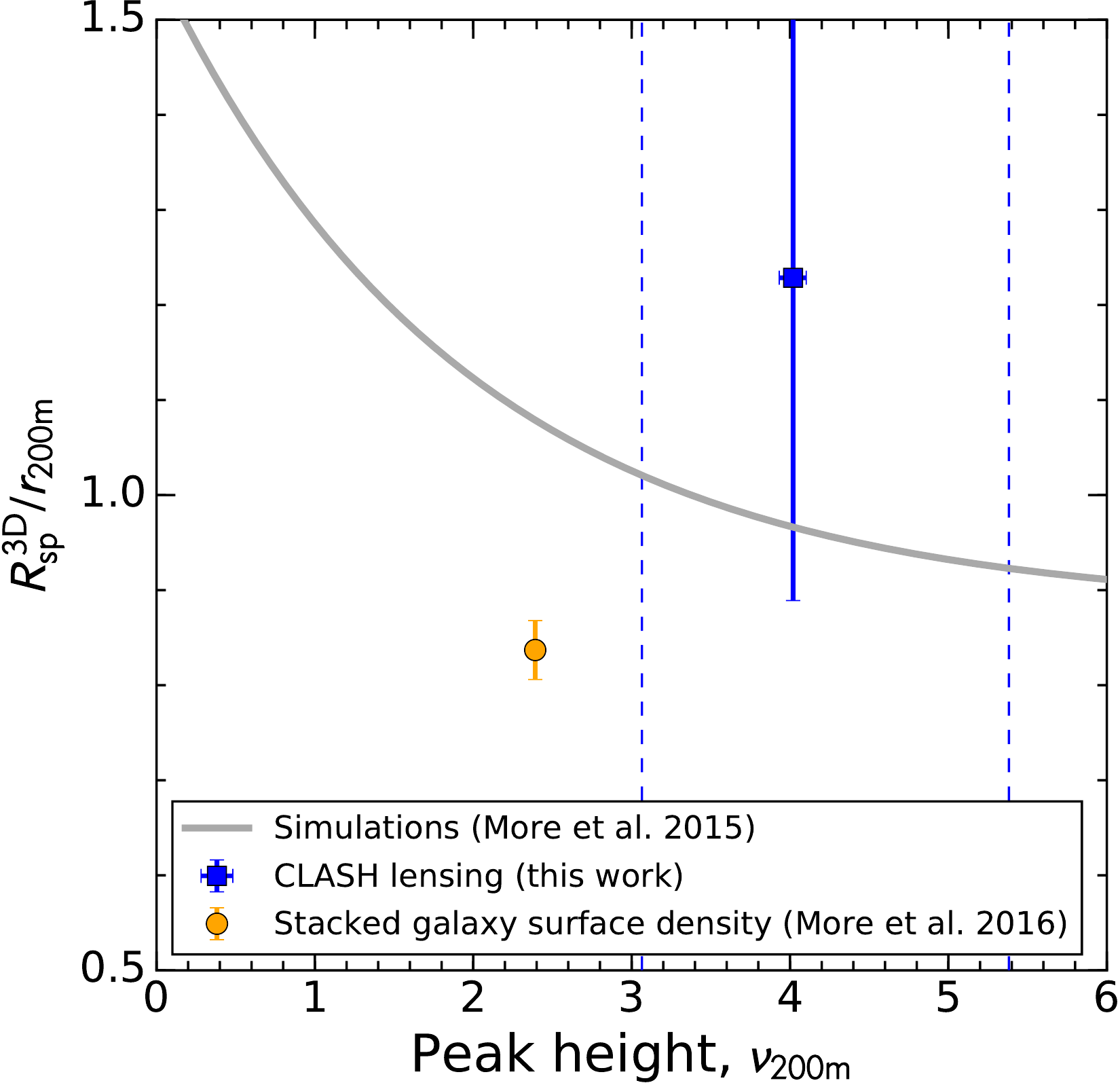}
  \end{array}
  $
 \end{center}
 \caption{
\label{fig:Rsp_Gamma}
Comparison of our CLASH lensing constraints on $\Rsp/r_\mathrm{200m}$
 against $\Lambda$CDM predictions at $z=0.337$ \citep[gray lines,][]{More2015splash}. 
Left panel: the relation between mass accretion rate and splashback
 radius. The blue horizontal line and the shaded area represent,
 respectively, the best-fit model value and the 68\% confidence interval 
 of the DK14 parameter $\Rsp/r_\mathrm{200m}$ inferred for the CLASH
 sample (Figure \ref{fig:posterior}). 
The CLASH lensing constraints on the splashback radius overlap with a
 broad representative range of mass accretion rates predicted for DM
 halos.  
 Right panel: the relation between peak height and splashback
 radius. Similarly, the CLASH results (blue square with error bars) are
 in agreement with the theoretical expectation 
 $\Rsp/r_\mathrm{200m}\simeq 0.97$ evaluated at the effective peak
 height of the CLASH sample, $\nu_\mathrm{200m} = 4.0\pm 0.1$. 
 The blue vertical dashed lines mark the range of $\nu_\mathrm{200m}$ peak
 heights covered by our sample.
 The observational results of \citet{More2016splash} for their full
 sample at $z=0.24$ are shown as a filled circle (see Section
 \ref{sec:discussion:obs}). Owing to the CLASH X-ray selection, there
 could be a bias toward higher values of $\Rsp/r_\mathrm{200m}$ for our
 sample.
} 
\end{figure*}

\subsection{Comparison with Simulation Results}
\label{sec:discussion:sims}

We begin by making sure that our results for the standard profile 
parameters, namely the concentration and Einasto shape parameters, are
congruent with expectations from simulations. The best-fit values for
concentration are relatively independent of the fitting function chosen
(Table \ref{tab:models}): the NFW fit results in
$c_\mathrm{200c} = 3.66$, the Einasto fit in
$c_\mathrm{200c} = 3.30$, and the DK14 fit in
$c_\mathrm{200c} = 3.58$. These values are in excellent agreement with
the model of \citet{Diemer+Kravtsov2015} which estimates the mean
concentration to be 
$c_\mathrm{200c} = 3.65$ for
$M_\mathrm{200c} = 10.11 \times 10^{14}\Msunh$,
$z=0.337$, and the cosmology assumed in this paper. The model of
\citet{Diemer+Kravtsov2015} is based on NFW concentrations,
whereas Einasto concentrations are expected to be about 10\% lower at
those masses and redshifts, in excellent agreement
with our results  
\citep[Figure 5 of \citealt{Dutton+Maccio2014},][]{Meneghetti2014clash,Sereno2016einasto}.
Our NFW constraints on the halo concentration
$c_\mathrm{200c}=3.66\pm 0.11$ (with 16 clusters) are also in agreement
with the expectations for the CLASH sample, namely a mean value of 3.87
and a standard deviation of 0.61, accounting for both selection and
projection effects \citep{Meneghetti2014clash}. 

Since the DK14 profile assumes an Einasto profile for the 1-halo term,
the Einasto parameters are of particular interest. The Einasto profile
varies in steepness with halo radius, at a rate given by a shape
parameter $\alpha$. \citet{Gao+2008} showed that this parameter is, to a 
good approximation, a function of only the peak height, $\nu$ \citep[see
also][]{Dutton+Maccio2014}. The CLASH mass corresponds to a peak
height of $\nu_{\rm 200c} = 3.76$ and thus $\alpha = 0.29$,
significantly higher than the values of about $0.2$ found in our
fits. While the results of neither \citet{Gao+2008} nor
\citet{Dutton+Maccio2014} are well constrained at such extreme peak  
heights, it is clear that $\alpha$ in their simulations exceeds $0.2$
significantly at high $\nu$. This tension already emerged as a
$\approx 1\sigma$ difference in the analysis of
U16. With the improved radial rescaling applied in
this paper, the significance of the difference increases to $3.5
\sigma$. 

We have tested whether the steepening due to the splashback radius
can bias the fitted $\alpha$ high compared to the value preferred
by the inner profile. We find that such a bias can indeed appear 
depending on the fitted radial range (a larger range leads to 
larger bias) and the weights given to each radial bin. However, 
\citet{Dutton+Maccio2014} fitted out to $1.2 \rvir$ and 
weighted the bins by the number of particles, and for these 
parameters the bias is smaller than $1\%$.
Another possible explanation is that the X-ray-selected CLASH
clusters are preferentially relaxed systems compared to the 
average population \citep{Meneghetti2014clash}. However, the
relation of
\citet{Gao+2008} and \citet{Dutton+Maccio2014} for $\alpha$ is
based on halo samples that exclude unrelaxed halos, although this
selection may not capture the entire effect present in the CLASH
sample. On the other hand,
taking into account baryonic effects in nonradiative hydrodynamical
$N$-body simulations, \citet{Meneghetti2014clash} find that the Einasto 
shape parameter for cluster-size halos
lies in the range $\alpha = 0.21 \pm 0.07$,
indicating  that our measurement is only moderately in tension with
simulations.

Finally, we compare our inferences regarding the splashback radius
$\Rsp$ with the simulation results of \citet{More2015splash} who
predicted that the ratio of $\Rsp$ and $r_\mathrm{200m}$ depends
primarily on the mass accretion rate of halos, with a less 
important dependence on redshift. In the left panel of Figure
\ref{fig:Rsp_Gamma}, we compare the ratio
$\Rsp/r_\mathrm{200m}=1.23^{+2.33}_{-0.34}$
(blue shaded band), inferred
for our sample at $z_\mathrm{l}^\mathrm{eff}=0.337$, with the mean
relation in simulations (gray band),   
\begin{equation}
\frac{\Rsp}{r_\mathrm{200m}} =
 0.58\left[1+0.63\Omega_\mathrm{m}(z)\right]
 \left(1+1.08 \exp\left[-\frac{\Gamma}{2.26}\right]\right),
\end{equation}
where $\Gamma\equiv d\ln{\Mvir}/d\ln{a}$ is the mass accretion rate
\citep{More2015splash}.
This comparison shows that the CLASH lensing constraints on the
splashback radius overlap with a broad range of mass accretion rates
$\Gamma$.
Our lower $1\sigma$ (16th percentile) limit of
$\Rsp/r_\mathrm{200m} \simgt 0.89$ translates into an upper limit on the accretion
rate of $\Gamma\simlt 4.0$. 
This limit is not particularly informative, since only a very small fraction
of halos experience such rapid accretion (DK14).  

Since we cannot directly measure the mass accretion rate of the
CLASH cluster sample, we cannot independently verify whether this
prediction is congruent with observations. For this reason,
\citet{More2015splash} also provide an approximate relation for the mean 
splashback radius as a function of peak height and redshift, 
\begin{equation}
\frac{\Rsp}{r_\mathrm{200m}} =
 0.81\left(
      1+0.97\exp
      \left[
       -\frac{\nu_\mathrm{200m}}{2.44}
      \right]
     \right) \,.
\end{equation}
The right panel of Figure \ref{fig:Rsp_Gamma} shows a comparison of this
fitting function with our results for the splashback radius and the peak
height of the CLASH cluster sample. Our inferred range of possible
splashback radii covers the entire range of values suggested by the
simulation results, showing that our results are generally compatible with
simulations. While the results of \citet{More2015splash} were based on
DM-only simulations, \citet{Lau2015} broadly confirmed the results of
DK14 in hydrodynamical simulations of individual clusters. Thus, we have
currently no reason to assume that baryons affect the location of the
splashback radius significantly.


\subsection{Projection Effects in Halo Profiles}
\label{sec:discussion:projection}

In the previous sections, we defined $\Rsp$ as the halo radius where the
logarithmic slope of the three-dimensional density profile is
steepest. However, the quantity actually measured is the two-dimensional
density profile, both when observing cluster member density profiles as
in \citet{More2016splash} or the lensing signal of clusters as in this
work. Thus, it is important to understand the relation between the
location of the radii of the steepest slope in thee dimensions and in
projection, $\Rsp$ and $R_\mathrm{sp}^\mathrm{2D}$. 

We have investigated this relation using the simulated halo sample
of \citet{More2015splash}. In all cases, the steepening of the
two-dimensional mass profile is less pronounced than that of the
three-dimensional profile, as highlighted in Figure 14 of DK14. At high
mass accretion rates ($\Gamma \geq 3$),
$R_\mathrm{sp}^\mathrm{2D} / \Rsp$ approaches a fixed ratio of about
$0.8$, in agreement with our measurements (Section \ref{subsec:Rsp}).  
At lower mass accretion rates, however, the $R_\mathrm{sp}^\mathrm{2D}$ 
derived from the profiles exhibits huge scatter and a seemingly random 
pattern.

The difficulty in deriving a valid $R_\mathrm{sp}^\mathrm{2D}$ from
projected measurements can be understood by considering a few
realizations of the DK14 profile with a power-law outer profile
representing the 2-halo term (see DK14 for details).
The location of the steepest slope in
three dimensions is a trade-off between the steepening 1-halo term and
the 2-halo term. The steepening is less pronounced for halos with
lower mass accretion rates. Thus, in projection, the 2-halo term has a
substantial impact on the apparent location of
$R_\mathrm{sp}^\mathrm{2D}$, which emerges at a much smaller radius that
is unrelated to the steepening term (Figure \ref{fig:Gamma3d}),  a
problem that becomes more serious at small peak heights.

These results highlight the importance of forward-modeling the effects
of the steepening based on the underlying three-dimensional density
profile, rather than attempting to derive $\Rsp$ from
$R_\mathrm{sp}^\mathrm{2D}$ directly (e.g., using Gaussian process
modeling).

\subsection{Compatibility with Measurements from Cluster Member Density Profiles}
\label{sec:discussion:obs}

The splashback radius was first unambiguously detected in observations by
\citet{More2016splash} who stacked surface number density profiles of
cluster member galaxies for a large number of clusters.
Their clusters were split into two subsamples with high and low
concentrations of member galaxies ($c_\mathrm{gal}$) at fixed richness
and redshift \citep{Miyatake2016bias}.
These high- and low-$c_\mathrm{gal}$ samples were expected to represent
populations of high and low $\Gamma$, respectively. However, \citet{Zu2016} 
have shown that the parameter $c_\mathrm{gal}$ used in \citet{More2016splash} 
is strongly contaminated by projection effects and is likely sensitive to 
the large-scale environment of the clusters rather than their internal structure.
We thus limit our comparison to the full sample of 
\citet{More2016splash} whose inferred splashback radius is
$\Rsp/r_\mathrm{200m}=0.837\pm 0.031$, with a weak-lensing mass of
$M_\mathrm{200m}\simeq 1.87\times 10^{14} \Msunh$ at $z=0.24$ (S. More
2016, private communication).

Figure \ref{fig:Rsp_Gamma} shows that our lower limit on $\Rsp$ is
higher than this value, but overlaps with the full-sample measurement 
at the $\sim 1\sigma$ level. We note that we expect the average mass 
accretion rate of the CLASH sample to be low due to a high fraction of
relaxed objects \citep{Meneghetti2014clash}. Hence, owing to the selection 
effects, there could be a bias toward higher values of
$\Rsp/r_\mathrm{200m}$ in our sample.

\section{Summary}
\label{sec:summary}

We have developed methods for modeling averaged cluster lensing   
profiles scaled to a chosen halo overdensity $\Delta$, which can be
optimized for the extraction of features that are local in radius,
in particular the steepening due to the splashback radius in the
outskirts of collisionless DM halos.  
We have examined the ensemble mass distribution of 16 CLASH
X-ray-selected clusters by forward-modeling the gravitational lensing
data obtained by \citet{Umetsu2016clash}. Our main conclusions are as
follows.
\begin{itemize}
\item Regardless of the overdensity chosen, the CLASH
ensemble mass profile in projection is remarkably well described by a
scaled NFW or Einasto density profile out to
$R\sim 1.2r_\mathrm{200m} \sim 2.5\Mpch$, beyond which the
data exhibit a flattening with respect to the NFW or Einasto profile. 
\item We constrain the NFW halo concentration to
      $c_\mathrm{200c}=3.66\pm 0.11$ at
      $M_\mathrm{200c}\simeq 1.0\times 10^{15}\Msunh$,
      consistent with previous work based on the same input data
      \citep{Umetsu2016clash}. Our new analysis using scaled profiles
      provides tighter constraints on the halo shape and structural
      parameters ($c$ and $\alpha$) than the conventional
      stacking.  
\item
     We do not find statistically significant evidence for the existence
     of the splashback radius in the CLASH lensing data.
     At the current sensitivity, this result is in line with expectations 
     from simulated, synthetic observations.
     Assuming the DK14 profile form and generic priors calibrated with
     simulations, we have placed a lower limit on the splashback radius
     of the clusters, if it exists, of
     $\Rsp/r_\mathrm{200m}>0.89$ or
     $\Rsp>1.83\Mpch$
     at 68\% confidence. This constraint is in agreement with 
     $\Lambda$CDM predictions.
\item The gradient feature in the outskirts is most pronounced for a
      scaling with $r_\mathrm{200m}$, consistent with simulation results of
      \citet{Diemer+Kravtsov2014} and \citet{Lau2015}.
\end{itemize}

The results obtained here are generally favorable in terms of the 
standard explanation for DM as effectively  collisionless and
nonrelativistic on sub-megaparsec scales and beyond, with an excellent match
between lensing data and $\Lambda$CDM predictions for high-mass
clusters. 
This study represents a first step toward using cluster gravitational
lensing to examine detailed predictions from collisionless $\Lambda$CDM
simulations regarding the shape and universality of the outer density 
profiles.
Such predictions can, in principle, be unambiguously tested across a
wide range of halo masses, redshifts, and accretion rates, with large
statistical samples of clusters from ongoing and planned lensing surveys
such as the Subaru Hyper Suprime-Cam survey \citep{Miyazaki2015}, the
Dark Energy Survey, and the {\em WFIRST} and {\em Euclid} missions. 


\acknowledgments
This work was made possible in part by the availability of high-quality
lensing data produced by the CLASH team. We are grateful to the CLASH
team who enabled us to carry out the work. We thank all authors of
\citet{Umetsu2014clash,Umetsu2016clash} and \citet{Zitrin2015clash} for 
their contributions to the lensing analyses used here.
We thank our referee for his valuable suggestions that
improved the paper significantly.
We thank Andrey Kravtsov and Surhud More for important suggestions
and detailed comments on a draft of this paper.
We acknowledge very fruitful discussions with
Nobuhiro Okabe,
Ho Seong Hwang,
Arman Shafieloo,
Zuhui Fan,
and
Congyao Zhang.
This work is partially supported by the Ministry of Science and
Technology of Taiwan under the grants 
MOST 103-2112-M-001-030-MY3
and
MOST 103-2112-M-001-003-MY3.

\clearpage 

\begin{appendix}

\section{Scaled DK14 Model}
\label{appendix:dk14}

We express the scaled DK14 density profile as
\begin{equation}
 f_{\Delta}(x) = f_\mathrm{inner}(x)\,f_\mathrm{trans}(x) + f_\mathrm{outer}(x)
\end{equation}
with
\begin{equation}
 \begin{aligned}
  f_\mathrm{inner}(x) &= \exp\left[ -\frac{2}{\alpha}c_\Delta^\alpha(x^\alpha-1) \right],\\ 
  f_\mathrm{trans}(x) &= \left[1+ \left(\frac{x}{\tau_\Delta}\right)^\beta\right]^{-\frac{\gamma}{\beta}},\\
  f_\mathrm{outer}(x) &= \frac{B_\Delta}{\epsilon_{\Delta} + x^{s_\mathrm{e}}},
 \end{aligned}
\end{equation}
where $c_\Delta=r_\Delta/r_\mathrm{s}$,
$\tau_\Delta=r_\mathrm{t}/r_\Delta$, 
$B_\Delta=b_\mathrm{e}(\rho_\mathrm{m}/\rho_\Delta)(5r_\mathrm{200m}/r_\Delta)^{s_\mathrm{e}}$
with $\rho_\Delta\equiv\rho_\mathrm{Einasto}(r_\Delta)=\rho_\mathrm{s}\exp\left[-(2/\alpha)(c_\Delta^\alpha-1)\right]$,
and
$\epsilon_\Delta=\Delta_\mathrm{max}^{-1}(5r_\mathrm{200m}/r_\Delta)^{s_\mathrm{e}}$.
The (unscaled) DK14 density profile is obtained as $\Delta\rho(r)=\rho_\Delta f_{\Delta}(r/r_\Delta)$. The derivatives of the inner, transition, and outer terms are 
\begin{equation}
 \begin{aligned}
  -x df_\mathrm{inner}/dx &= 2c_\Delta^\alpha x^\alpha f_\mathrm{inner},\\
  -x df_\mathrm{trans}/dx &= \gamma\frac{(x/\tau_\Delta)^\beta}{1+(x/\tau_\Delta)^\beta}f_\mathrm{trans},\\
  -x df_\mathrm{outer}/dx &=s_\mathrm{e} \frac{x^{s_\mathrm{e}}}{\epsilon_\Delta+x^{s_\mathrm{e}}}f_\mathrm{outer}. 
 \end{aligned}
\end{equation}
The logarithmic gradient of the DK14 density profile $\Delta\rho(r)$
with $r=r_\Delta x$ is thus given by
\begin{eqnarray}
\frac{d\ln\Delta\rho}{d\ln{r}}
=\frac{d\ln{f_{\Delta}}}{d\ln{x}}
&=& \left(
x\frac{df_\mathrm{inner}}{dx} f_\mathrm{trans} +
  xf_\mathrm{inner}\frac{df_\mathrm{trans}}{dx} +
  x\frac{df_\mathrm{outer}}{dx}
\right) \Big/ f_{\Delta}\\
&=&
-\left[
2(c_\Delta x)^\alpha f_\mathrm{inner}f_\mathrm{trans}
+
 \gamma\frac{(x/\tau_\Delta)^\beta}{1+(x/\tau_\Delta)^\beta}
 f_\mathrm{inner}f_\mathrm{trans}
+
s_\mathrm{e} \frac{x^{s_\mathrm{e}}}{\epsilon_\Delta+x^{s_\mathrm{e}}}f_\mathrm{outer}
\right] \Big{/} f_{\Delta}.
\end{eqnarray}
For the Einasto model ($f_\mathrm{trans}=1$, $f_\mathrm{outer}=0$),
$d\ln{\Delta\rho}/d\ln{r}=-2(c_\Delta x)^\alpha=-2(r/r_\mathrm{s})^\alpha$.


In this work, the splashback radius $\Rsp$ is
defined as the location of the steepest slope of the three-dimensional
mass distribution $\Delta\rho(r)$. 
For a given set of the DK14 model parameters, we find the scaled
splashback radius $x_\mathrm{sp}\equiv\Rsp/r_\Delta$ from
\begin{equation}
\label{eq:Rsp}
x_\mathrm{sp} = \argmin_x  \frac{d\ln{f_{\Delta}}}{d\ln{x}}.
\end{equation}

The ratio of the splashback mass
$M_\mathrm{sp}=M(<\Rsp)$
to the overdensity mass $M_\Delta=M(<r_\Delta)$ is given by
\begin{equation}
 \label{eq:Msp}
 \frac{M_\mathrm{sp}^\mathrm{3D}}{M_\Delta}=\frac{\int_0^{x_\mathrm{sp}}\!dx\,x^2f_{\Delta}(x)}{\int_0^1\!dx\,x^2f_{\Delta}(x)}.
\end{equation}


\section{Synthetic Weak-lensing Data}
\label{appendix:mock}

We create a total of 50 source realizations of synthetic shear and
magnification catalogs for our sample of 16 CLASH clusters, each of 
which is modeled as a spherical NFW halo specified by its redshift
$z_\mathrm{l}$ (Table \ref{tab:sample}) and its parameters
$M_\mathrm{200c}$ and $c_\mathrm{200c}$, which were fixed to the observed 
central values (Table 2 of U16).

For each NFW cluster, we consider two configurations of the outer density 
profile, one with and one without a splashback-like feature. For the latter, 
the total density profile is given by a single NFW profile,
$\Delta\rho(r) = \rho_\mathrm{NFW}(r|M_\mathrm{200c},c_\mathrm{200c},z)$.
For the former, we employ a composite lens model that
produces an approximate splashback feature. We cannot use the DK14 profile directly
because it is not implemented in the \textsc{glafic} software used to perform ray-tracing simulations 
as described below. However, the code does implement a steepening 1-halo term
given by the truncated NFW profile of \citet[][BMO]{BMO}, $\rho_\mathrm{BMO}(r)$.
We approximate the 2-halo term by a softened isothermal (SI) profile,
$\rho_\mathrm{SI}(r)$: 
$\Delta\rho=\rho_\mathrm{BMO} + \rho_\mathrm{SI}$.
The BMO density profile is expressed as
$\rho_\mathrm{BMO}(r)=\rho_\mathrm{NFW}(r|M_\mathrm{200c},c_\mathrm{200c},z)\times f_\mathrm{trans}(r|\beta,\gamma,r_\mathrm{t})$
with $\beta=2$ and $\gamma=4$.
The truncation radius $r_\mathrm{t}$ is set to
$r_\mathrm{t}=1.1\rvir\approx r_\mathrm{200m}$
We take the outer SI profile to be
$\rho_\mathrm{SI}(r)=\rho_\mathrm{c}/[1+(r/r_\mathrm{c})^2]$
with
$\rho_\mathrm{c}=1.5\times 10^{12}h^2\,M_\odot$\,Mpc$^{-3}$
and
$r_\mathrm{c}=2.2\Mpch$,
so as to give a splashback feature at
$\Rsp/r_\mathrm{200m}\sim 0.9$ for our CLASH sample.
Here, the normalization of the SI profile is chosen to be three times
lower than that of the 2-halo term in projection
\citep[][see their Figure 7]{Umetsu2014clash},
because the standard halo model
($\Delta\rho =\rho_\mathrm{BMO}+\rho_\mathrm{2h}$), by design, does not
produce a steepening relative to the NFW profile 
\citep[][see their Figure 2]{Oguri+Hamana2011}.

In general, source galaxy catalogs used for the shear and the
magnification analysis are different because we apply different size,
magnitude, and color cuts in source selection for measuring the
shear and the magnification effects \citep{Umetsu2014clash}.
We assume, for simplicity, that the two galaxy samples are identical.
We ignore the cosmic noise contribution from projected
uncorrelated large-scale structures here because it is subdominant in
the total error budget (see Figure 1 of U16).
For all clusters, we assume the same survey parameters and source
properties as described below.

To produce synthetic magnification-bias data sets, we perform
ray-tracing simulations
with both NFW and BMO+SI lenses
\citep{2000ApJ...534...34W,BMO,Oguri+Hamana2011}
using the public package \textsc{glafic} \citep{glafic}.
We assume a maximally depleted sample of sources with
$s\equiv d\log_{10}N(<m)/dm=0$, for which the effect of 
magnification bias is purely geometric\footnote{  
For a depleted population of sources with
$s<0.4$, the net effect of magnification bias is 
dominated by the geometric area distortion \citep{Umetsu2013}, and is
insensitive to the intrinsic source luminosity function. This is the
case for the $BR_\mathrm{C}z'$-selected red galaxy samples with
$\langle s\rangle \sim 0.15$
used for the magnification bias measurements of
\citet{Umetsu2014clash}.}
and their lensed source counts can be inferred from the lensed image 
positions.
%
We randomly distribute $N_\mathrm{s}=14,336$ source galaxies over an area
of $32\times 32$\,arcmin$^2$ in the source plane centered on the cluster.
This corresponds to an unlensed source density of 
$\overline{n}_\mathrm{s}=14$ galaxies per arcmin$^{2}$, 
matched to the typical (median) value found in the CLASH weak-lensing
observations of \citet[][their Table 4]{Umetsu2014clash}.
The source plane is placed at a redshift of $z_\mathrm{s}=1.0$, the
median depth of their magnification samples \citep{Umetsu2014clash}.   

For creation of synthetic shear catalogs, we draw $N_\mathrm{s}$ random
source ellipticities from the Gaussian intrinsic ellipticity
distribution given by Equation (12) of \citet{Schneider+2000}, with the
rms intrinsic ellipticity assumed to be $0.3$.
For each galaxy, we transform the source ellipticity into the
image ellipticity at the image position according to Equation (1) of
\citet{Schneider+2000}. 

Our simulations include the following major steps of data analysis,
signal reconstruction, and modeling processes:  
\begin{enumerate}
 \item Measurements of the reduced tangential shear and magnification
       bias profiles in $N_\mathrm{WL}=10$ log-spaced radial bins over the radial range
       $\theta\in [0.9\arcmin,16\arcmin]$ (Section \ref{subsec:data})
       from the respective input source catalogs, following the analysis
       procedures described in \citet{Umetsu2014clash} and U16.
 \item Reconstructions of the projected mass profile ($\bSigma$) from the binned
       shear and magnification constraints obtained in the first step. 
 \item Ensemble characterization of the cluster $\bSigma$ profiles using
       the scaled DK14 model and the priors described in Sections
       \ref{subsec:DK14} and \ref{subsec:bayesian}. 
\end{enumerate}
For the second step, we use the cluster lensing mass inversion
(\textsc{clumi}) code \citep{Umetsu+2011,Umetsu2013} 
as implemented in U16 but without using inner strong-lensing
constraints, and assume perfect knowledge of the source properties,
namely
$z_\mathrm{s}=1.0$,
$\overline{n}_\mathrm{s}=14$ galaxies arcmin$^{-2}$,
and $s=0$.
Otherwise,
synthetic data are processed in the same manner as the CLASH data
described in Section \ref{subsec:data}.
In the third step, we find, for each source realization, the global
maximum a posteriori 
estimate of the joint posterior distribution to infer the best-fit 
DK14 parameters.
 
\end{appendix}



\end{document}